\documentclass[prl,aps,superscriptaddress,floatfix,tightenlines,showpacs,twocolumn]{revtex4-2}
\usepackage{graphicx}
\usepackage{dcolumn}   
\usepackage{bm}        
\usepackage{amssymb}   
\usepackage{amsmath}
\usepackage{url}
\usepackage{layout}
\usepackage{color}
\usepackage{epsfig}
\usepackage{graphicx}
\usepackage{mathrsfs}
\usepackage{bm}
\usepackage{appendix}
\usepackage{color}
\usepackage[thinlines]{easytable}
\usepackage{makecell}
\usepackage{tikz,pgf}
\usepackage{booktabs}
\usepackage{float}
\usepackage{hyperref}
\hypersetup{colorlinks,allcolors=blue}

\begin{document}

\title{Impact of Pressure and Apical Oxygen Vacancies on Superconductivity in La$_3$Ni$_2$O$_7$
}
\author{Chen Lu}
\thanks{These authors contributed equally.}
\affiliation{School of Physics, Hangzhou Normal University, Hangzhou 311121, China}
\author{Ming Zhang}
\thanks{These authors contributed equally.}
\affiliation{Zhejiang Key Laboratory of Quantum State Control and Optical Field Manipulation, Department of Physics, Zhejiang Sci-Tech University, 310018 Hangzhou, China}
\author{Zhiming Pan}
\thanks{These authors contributed equally.}
\affiliation{Department of Physics, Xiamen University, Xiamen 361005, Fujian, China}
\author{Congjun Wu}
\affiliation{New Cornerstone Science Laboratory, Department of Physics, School of Science, Westlake University, Hangzhou 310024, Zhejiang, China}
\affiliation{Institute for Theoretical Sciences, Westlake University, Hangzhou 310024, Zhejiang, China}
\affiliation{Key Laboratory for Quantum Materials of Zhejiang Province, School of Science, Westlake University, Hangzhou 310024, Zhejiang, China}
\affiliation{Institute of Natural Sciences, Westlake Institute for Advanced Study, Hangzhou 310024, Zhejiang, China}
\author{Fan Yang}
\email{yangfan\_blg@bit.edu.cn}
\affiliation{School of Physics, Beijing Institute of Technology, Beijing 100081, China}

\begin{abstract}
\noindent The bilayer nickelate La$_3$Ni$_2$O$_7$ under pressure has recently emerged as a promising system for high-$T_c$ superconductivity.
In this work, we investigate the fate of the superconducting properties in La$_3$Ni$_2$O$_{7}$ under pressure, focusing on the effects of structural deformation and apical oxygen vacancies. 
Employing a low-energy effective $t$-$J_{\parallel}$-$J_{\perp}$ model for the $3d_{x^2-y^2}$ orbitals within the slave-boson mean-field approach, 
we demonstrate that the pairing strength is significantly enhanced in the high-pressure tetragonal $I4/mmm$ phase compared to the ambient pressure orthorhombic $Amam$ phase. 
Furthermore, by simulating random configurations of apical oxygen vacancies, we show that oxygen vacancies suppress both pairing strength and superfluid density. 
These results underscore the critical role of pressure and oxygen stoichiometry in tuning the SC of La$_3$Ni$_2$O$_7$, providing key insights into optimizing its high-$T_c$ behavior.
\end{abstract}
\maketitle

\noindent{\bf Introduction}

Ruddlesden-Popper (RP) phase nickelates series La$_{n+1}$Ni$_n$O$_{3n+1}$ have been a subject of sustained and intense research due to their potential for realizing high-$T_c$ superconductivity (SC)  \cite{taniguchi1995transport,seo1996electronic,kobayashi1996transport,greenblatt1997ruddlesden,greenblatt1997electronic,ling2000neutron,wu2001magnetic,fukamachi2001nmr,voronin2001neutron,Bannikov2006,hosoya2008pressure,pardo2011dft,nakata2017finite,mochizuki2018strain,li2020epitaxial,song2020structure,barone2021improved,Wang2022LNO}. 
In a recent pivotal development, SC phenomena exhibiting a critical temperature $T_c$ of approximately $80$K was discovered in La$_3$Ni$_2$O$_7$ ($n=2$) under a pressure exceeding 14 GPa \cite{Wang2023LNO,Wang2023LNOb,YuanHQ2023LNO,wang2023LNOpoly,wang2023la2prnio7,wang2024bulk}, 
attracting great interests ~\cite{liu2024electronic,yang2024orbital,zhang2023pressure,wang2023structure,zhou2023investigations,cui2024strain,dan2024spin,abadi2024electronic,li2024ultrafast,wang2024bulk,li2024electronic,zhang2024doping,ren2025resolving,li2024distinguishing,zhou2024revealing,su2024strongly,mijit2024local,chen2024unveiling,Dong2024vis,shi2025prerequisite,li2025ambient,huo2025low,zhang2025damage,YaoDX2023,WangQH2023,YangF2023,lechermann2023,Kuroki2023,HuJP2023,zhang2023structural,lu2023bilayertJ,oh2023type2,liao2023electron,qu2023bilayer,Yi_Feng2023,jiang2023high,zhang2023trends,qin2023high,tian2023correlation,jiang2023pressure,lu2023sc,kitamine2023,luo2023high,zhang2023strong,pan2023rno,lange2023mixedtj,yang2023strong,lange2023feshbach,kaneko2023pair,qu2023roles,kakoi2023pair,fan2023sc,wu2024deconfined,zhang2024electronic,ryee2024quenched,ni2024spin,Lu2024interplay,Ouyang2024absence,liu2024origin,chen2024electronic,li2024pressure,xie2024strong,yue2025correlated,wang2025mottness}. 
Furthermore, SC with a $T_c\approx 40$K was realized in La$_3$Ni$_2$O$_7$ thin films under ambient pressure \cite{ko2024signatures,zhou2024ambient}, representing a crucial step forward towards practical applications and experimental investigations ~\cite{li2025photoemission,bhatt2025resolving,liu2025superconductivity,zhong2025epitaxial}. 
Additionally, experimental evidence of SC with a $T_c \approx 20$K in another RP-phase material, La$_4$Ni$_3$O$_{10}$ ($n=3$) \cite{li2023trilayer,zhu2023trilayer,zhang2023trilayer}, has also attracted significant attention \cite{Yuan2024la3,sakakibara2023trilayer,li2024la3,kakoi2023multiband,leonov2024la3,tian2024effective,wang2024nonfermi,labollita2024electronic,zhang2024prediction,yang2024effective,luo2024trilayer,zhang2024s,li2024ultrafast,lechermann2024electronic,huang2024interlayer,oh2024type,qin2024frustrated,xu2024origin,du2024correlated,huo2024electronic,yang2024decom,zhang2024magnetic,huang2024signature,liu2024evolution,deswal2024dynamics,zhao2024electronic}.
A recent discovery of SC was found in {L}a$_5${N}i$_3${O}$_{11}$ under pressure with $T_c\approx 60$ K \cite{shi2025superconductivity}.
These advancements underscore the important role of this nickelate series as a crucial platform for exploring high-$T_c$ SC.

The phase diagram of bilayer La$_3$Ni$_2$O$_{7}$ has been explicitly explored across a wide range of temperatures and pressures ~\cite{Wang2023LNO,li2024pressure}, 
revealing a complex and rich phenomenology. 
In particular, the complex nature of the phase diagram and the potential for unconventional SC have attracted a subsequent surge of theoretical investigations aimed at revealing the underlying SC mechanism \cite{YaoDX2023,WangQH2023,YangF2023,lechermann2023,Kuroki2023,HuJP2023,zhang2023structural,lu2023bilayertJ,oh2023type2,liao2023electron,qu2023bilayer,Yi_Feng2023,jiang2023high,zhang2023trends,qin2023high,tian2023correlation,jiang2023pressure,lu2023sc,kitamine2023,luo2023high,zhang2023strong,pan2023rno,lange2023mixedtj,yang2023strong,lange2023feshbach,kaneko2023pair,qu2023roles,kakoi2023pair,fan2023sc,wu2024deconfined,zhang2024electronic,ryee2024quenched,ni2024spin,Lu2024interplay,Ouyang2024absence,liu2024origin,yue2025correlated,wang2025mottness}. 
Despite these intensive theoretical efforts, the precise nature of the pairing mechanism and symmetry in La$_3$Ni$_2$O$_{7}$ remains an open question and a subject of ongoing debate.
Two key factors, pressure and oxygen stoichiometry, play important role and complicate the problem.

Experimental studies have definitively demonstrated that the emergence of SC in La$_3$Ni$_2$O$_{7}$ is strongly and sensitively influenced by the application of external pressure \cite{Wang2023LNO,Wang2023LNOb,YuanHQ2023LNO,wang2023LNOpoly,wang2023la2prnio7,wang2024bulk,zhang2023pressure,wang2023structure,zhou2023investigations,huo2025low}.
 At ambient pressure (AP), the material adopts an orthorhombic $Amam$ crystal structure.
High-quality single crystals are metallic down to low temperatures, while oxygen deficiency can induce weakly insulating behavior. 
Under increasing pressure, La$_3$Ni$_2$O$_{7}$ undergoes sequential structural phase transitions \cite{Wang2023LNO,li2024pressure}.
Initially, it transforms from the $Amam$ phase to an orthorhombic $Fmmm$ phase around 14 GPa, 
accompanied by inter-bilayer Ni-O-Ni bond angle from approximately $168^{\circ}$ towards $180^{\circ}$, 
which significantly enhances inter-bilayer coupling.  
Subsequently, at high pressure (HP), it transforms into a tetragonal $I4/mmm$ phase. 
SC is presumed to emerge near the first structural phase transition into the $Fmmm$ phase, with the superconducting transition temperature $T_c$ subsequently decreasing as pressure increases \cite{Wang2023LNO,li2024pressure}.

The realization of SC is also highly sensitive to oxygen stoichiometry ~\cite{li2024pressure,ko2024signatures,liu2025superconductivity,zhou2023investigations,ren2025resolving,Dong2024vis,shi2025prerequisite,zhang2025damage}, emphasizing that sufficient oxygen content is crucial for stabilizing robust SC \cite{YangF2023,zhang2023la3ni2o6,wang2025mottness}.
Oxygen vacancies significantly alter the electronic structure and orbital interactions critical to SC. 
Experimental observations, including scanning transmission electron microscopy (STEM) and X-ray diffraction (XRD) ~\cite{Dong2024vis,li2024pressure}, have provided direct visualization of oxygen vacancies, particularly attributed to apical oxygen sites.
Transport measurements have revealed that oxygen vacancies can reduce the superconducting volume fraction and, in some samples, induce a filamentary SC state ~\cite{ko2024signatures}. 
Additionally, ozone annealing has been employed to manipulate oxygen stoichiometry, demonstrating the importance of oxygen control in optimizing SC ~\cite{liu2025superconductivity}.

In this paper, we employ strong correlated analysis to investigate the effects of pressure and random apical oxygen vacancies on the superconducting properties of La$_3$Ni$_2$O$_{7}$. 
To gain insight into the pairing mechanism, we focus on a low-energy effective $t$-$J_{\parallel}$-$J_{\perp}$ model \cite{lu2023bilayertJ} for the $3d_{x^2-y^2}$ orbital. We consider two representative regimes: 
AP $Amam$ phase and HP $I_4/mmm$ phase at $30$ GPa. 
At $30$ GPa, any intralayer anisotropy between the $x$ and $y$ directions is typically minimal. 
Therefore, for simplicity and to effectively capture the near-tetragonal high-pressure environment, we adopt the $I4/mmm$ phase with equivalent hopping integrals along these in-plane directions.
Our numerical results reveal a superconducting state in the AP $Amam$ phase, while the pairing strength is significantly weaker than that in the HP $I_4/mmm$ phase.
Furthermore, we simulate the influence of random oxygen vacancy configurations, demonstrating that apical oxygen vacancies act to suppress both pairing strength and superfluid density. 
These findings provide valuable theoretical insights into the impact of pressure and oxygen vacancies on the superconducting properties of La$_3$Ni$_2$O$_{7}$, offering guidance for future experimental and theoretical investigations.


\vspace{0.5\baselineskip}
\noindent{\bf Results}

\noindent{\bf Effective bilayer single-orbital model}

\begin{figure}[t!]
\centering
\includegraphics[width=0.45\textwidth]{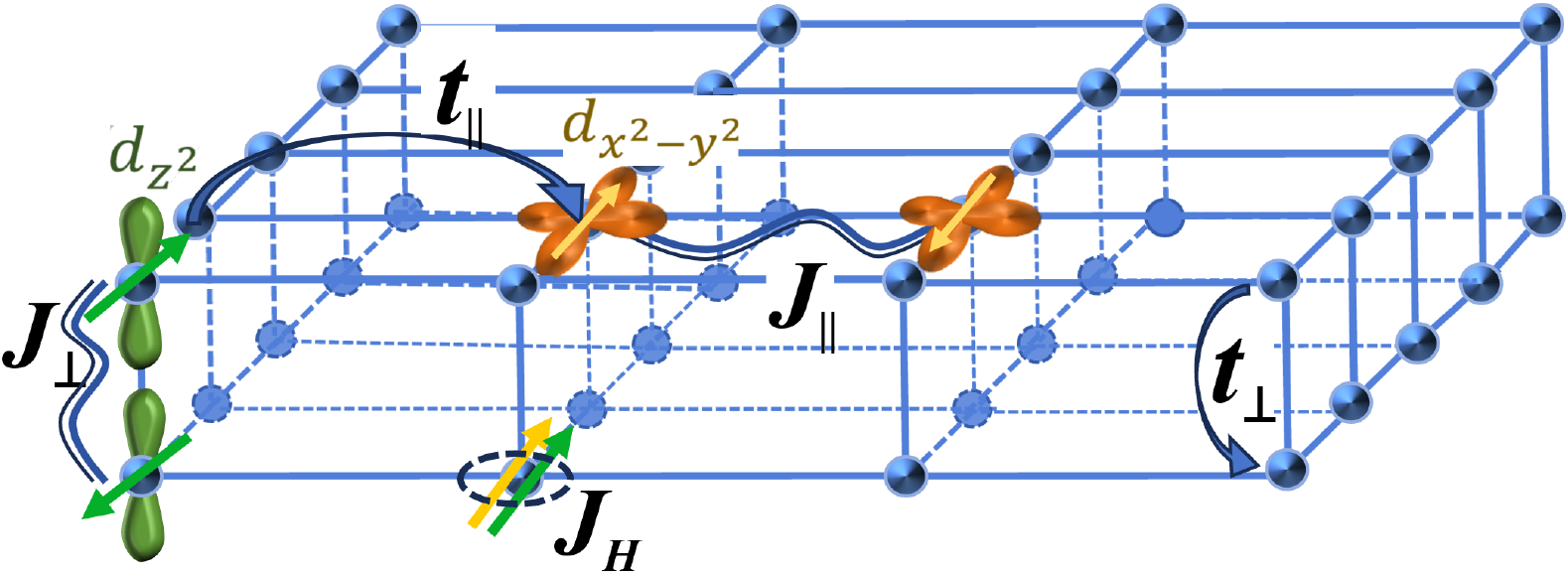}
\caption{Schematic illustration of the two-orbital bilayer $t$-$J$-$J_H$ model. 
The charge carriers primarily reside in the $3d_{x^2-y^2}$ orbitals, as indicated with wavy lines on the lattice, while the $3d_{z^2}$ orbitals are depicted as localized spins due to their single occupancy.
These localized spins interact via interlayer spin exchange $J_{\perp}$, reflecting the strong correlations present in the system. 
The onsite $E_g$ orbitals tend to form a spin-triplet configuration driven by Hund's rule coupling $J_H$.}
\label{fig1}
\end{figure}

The electronic properties of La$_3$Ni$_2$O$_{7}$ is dominated by the two $E_g$ orbitals \cite{YaoDX2023} within Ni$^{2.5+}$ ions.
Direct valence counting shows that the $3d_{z^2}$ orbital is nearly half-filled, while the $3d_{x^2-y^2}$ orbital is approximately quarter-filled. 
Besides, La$_3$Ni$_2$O$_{7}$ exhibits strong electronic correlations, as evidenced by optical measurements showing the suppression of electron kinetic energy ~\cite{liu2024electronic} and band renormalization revealed from angle-resolved photoemission spectroscopy (ARPES) data \cite{yang2024orbital}, placing the system proximity to a Mott regime. 
Under strong Hubbard repulsion, $3d_{z^2}$-orbital electrons become effectively localized \cite{yang2024orbital,YaoDX2023}. 
SC primarily contributes from the $3d_{x^2-y^2}$ orbital, which shares similarities with overdoped cuprates. 
However, the in-plane superexchange interaction between the $3d_{x^2-y^2}$ electrons alone is not sufficient to account for high-$T_c$ SC.

A critical aspect of the pairing mechanism in La$_3$Ni$_2$O$_{7}$ lies in the interplay between the $3d_{z^2}$ and $3d_{x^2-y^2}$ orbitals.  The corresponding effective $t$-$J$-$J_H$ model that captures this orbital entanglement is illustrated in Fig.~\ref{fig1}. The strong interlayer antiferromagnetic (AFM) superexchange associated with the $3d_{z^2}$ spins is transmitted to the $3d_{x^2-y^2}$ orbital via Hund’s coupling ~\cite{lu2023bilayertJ}. 
This mediates an effective interlayer spin exchange $J_{\perp}$ between $3d_{x^2-y^2}$ spins, which is essential for the emergence of SC in this system. 
The electronic properties of the $3d_{x^2-y^2}$ orbital under strong Hund's coupling can be captured by a bilayer $t$-$J_{\parallel}$-$J_{\perp}$ model ~\cite{lu2023bilayertJ,qu2023bilayer,oh2023type2}.

The minimal effective model consists of a bilayer $t$-$J_{\parallel}$-$J_{\perp}$ Hamiltonian can be expressed as follows:
\begin{equation}
\begin{aligned}
H=& \sum_{\langle i,j\rangle \alpha\sigma} -t_{\langle i,j\rangle}^{\parallel}\big(c_{x^2i\alpha\sigma}^{\dagger} c_{x^2j\alpha\sigma} +\text{h.c.}\big) 
\\
&-t_{\perp} \sum_{i\sigma} \big(c_{x^2i1\sigma}^{\dagger} c_{x^2i2\sigma} +\text{h.c.}\big) 
\\
&+\sum_{\langle i,j\rangle \alpha} J_{\parallel} \bm{S}_{x^2i\alpha} \cdot \bm{S}_{x^2j\alpha} 
+J_{\perp} \sum_{i} \bm{S}_{x^2i1} \cdot \bm{S}_{x^2i2},
\end{aligned}
\label{eq:x2-tJ-H}
\end{equation}
where $c_{x^2i\alpha\sigma}^{\dagger}$ is the electron creation operator for $3d_{x^2-y^2}$ orbital at the lattice site $i$;
$\alpha=1,2$ is the layer index for the NiO$_2$ plane and $\sigma=\uparrow,\downarrow$ is the spin index;
$\bm{S}_{x^2i\alpha}=\frac{1}{2}c_{x^2i\alpha}^{\dagger}[\bm{\sigma}]c_{x^2i\alpha}$ is the spin operator for $3d_{x^2-y^2}$ electron, with Pauli matrices $\bm{\sigma}=(\sigma_x,\sigma_y,\sigma_z)$.
The summation $\sum_{\langle i,j\rangle}$ takes over all the nearest-neighboring (NN) pair $\langle i,j\rangle$ within the NiO$_2$ plane. 
In the following study, $t$ is set as the energy unit. 
The super-exchange interaction is effectively approximately as $J_{\parallel}=4t_{\parallel}^2/U$, where $U$ is the Hubbard interaction in the $3d_{x^2-y^2}$ orbital. We consider the hole picture, where the hole density is given by $\delta_{h}=1-2x$.
Here, $x$ represents the electron filling for the $d_{x^2-y^2}$-orbital, with $x=0.25$ indicating quarter-filling.

This effective $3d_{x^2-y^2}$ orbital $t$-$J_{\parallel}$-$J_{\perp}$ model could potentially exhibit high-$T_c$ SC with interlayer $s$-wave pairing \cite{lu2023bilayertJ,qu2023bilayer,lange2023feshbach,lange2023mixedtj,zhang2023strong}.
The important role of Hund's coupling in stabilizing SC has also been numerically confirmed \cite{qu2023roles,kakoi2023pair}. The strong interlayer AFM exchange, crucial for SC, has been experimentally revealed by spin excitation studies, including resonant inelastic X-ray scattering \cite{chen2024electronic,zhong2025epitaxial} and inelastic neutron scattering \cite{xie2024strong}. 
These studies indicate a strong interlayer value for $SJ_{\perp}\approx 60-70$ meV, while the intralayer one is much weaker.
We further predict element substitution could enhance the SC based on this pairing mechanism \cite{pan2023rno}, a trend that is qualitatively consistent with recent experiments on La$_2$ReNi$_2$O$_7$ (Re:rare earth) \cite{wang2024bulk,li2025ambient}.

\vspace{0.5\baselineskip}
\noindent{\bf Pressure and Superconductivity}

\begin{figure}[t!]
\centering
\includegraphics[width=0.45\textwidth]{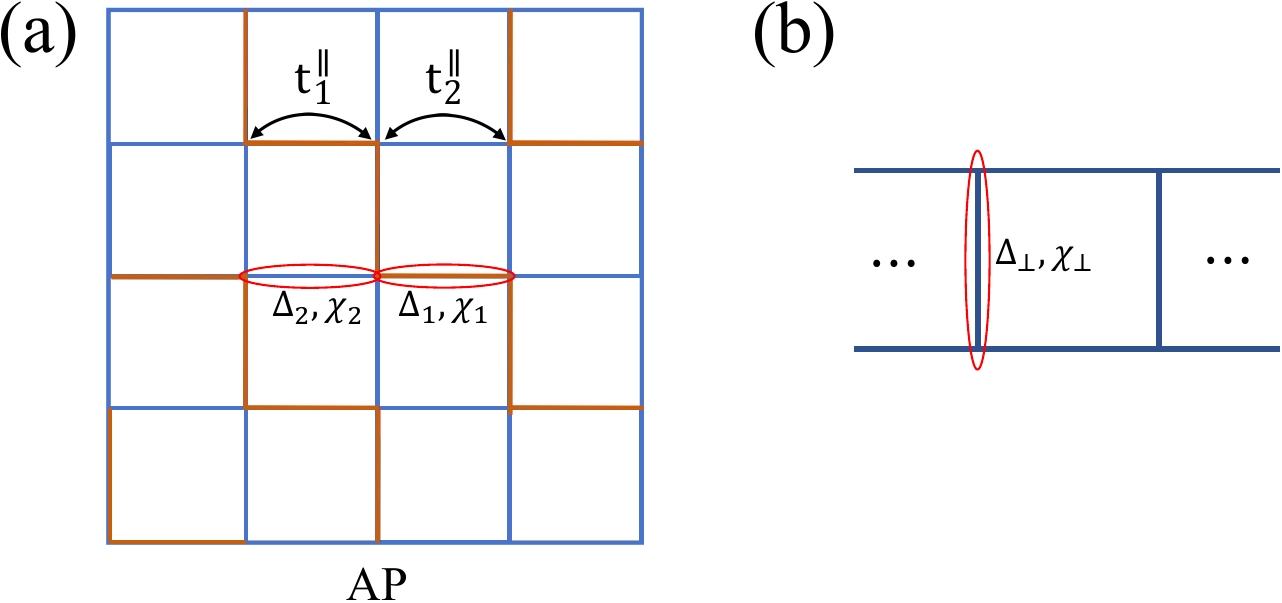}
\caption{Schematic diagram for the hopping process and order parameters. (a) Schematic diagram for the intra-layer nearest-neighboring hopping process at ambient pressure, and the two types of intra-layer order parameters($\Delta_{1,2}, \chi_{1,2}$). 
(b) Schematic diagram for the inter-layer order parameters ($\Delta_{\perp}, \chi_{\perp}$).}
\label{fig2}
\end{figure}

Pressure induces changes in the lattice structure, consequently modifying the underlying electronic properties.
Fig.~\ref{fig2}(a) and (b) illustrate the intra-layer and inter-layer hopping parameters, respectively, denoted as $t_{\langle i,j\rangle}^{\parallel}$.
In the AP phase, spatial inversion symmetry ensures that inter-layer order parameters are identical for symmetry-equivalent positions. 
Density functional theory (DFT) calculations \cite{liu2024origin} provide the AP and HP hopping parameters as shown in Table~\ref{tab:HP}. 
In the strong coupling limit, the spin superexchange interaction $J$ scales proportionally to $t^2$. 
Consequently, $J$ is expected to become stronger under pressure due to the increased hopping parameters.

\begin{table}[t]
	\centering
	\begin{TAB}(r,0.05cm,0.1cm)[2pt]{|c|c|c|c|}{|c|c|c|c|c|}
		hopping &  AP value (eV)& hopping & HP value (eV) \\
		$t_{1}^{\parallel}$  & $0.425$
		& $t^{\parallel}$ & $0.49$ \\
		$t_{2}^{\parallel}$  & $0.3976$
		& & \\
		$t_{xx}^{\perp}$  & $1.82\times 10^{-2}$
		& $t_{xx}^{\perp}$& $2.45\times 10^{-2}$  \\
		$t_{zz}^{\perp}$  & $0.597$
		& $t_{zz}^{\perp}$& $0.67$  \\
	\end{TAB}
	\caption{Table of the hopping parameters at ambient pressure (AP) and high pressure (HP).}
	\label{tab:HP}
\end{table}

\begin{table}[t]
	\centering
	\begin{TAB}(r,0.05cm,0.1cm)[2pt]{|c|c|c|c|c|}{|c|c|c|c|c|c|c|c|c|}
		AP OP &  GS value (eV) & & 
		HP OP &  GS value (eV) \\
		$\Delta_{\perp}$  & $-2.40\times 10^{-3}$
		& & $\Delta_{\perp}$ &$-1.38\times 10^{-2}$ \\
		$\Delta_{1}$  & $7.18\times 10^{-6}$
		& & $\Delta_{\parallel}$ & $6.83\times 10^{-5}$ \\
		$\Delta_{2}$  & $5.83\times 10^{-6}$
		& &  &  \\
		
		$\chi_{\perp}$ & $1.59\times 10^{-3}$
		& & $\chi_{\perp}$ & $2.54\times 10^{-3}$ \\
		$\chi_{1}$ & $1.97\times 10^{-2}$
		& & $\chi_{\parallel}$ & $3.23\times 10^{-2}$ \\
		$\chi_{2}$ & $1.96\times 10^{-2}$
		& &  &  \\
		
		&  filling
		& &  & filling \\
		$x$ &  $0.25$
		& & $x$ & $0.3$ \\
	\end{TAB}
	\caption{Table of the ambient pressure (AP) and high pressure (HP) order parameters (OPs) as well as the physical electron filling $x$. The OPs are calculated by the slave-boson mean-field theory in the ground state (GS).}
	\label{tab:MForderParameter}
\end{table}

\begin{figure}[t!]
\centering
\includegraphics[width=0.48\textwidth]{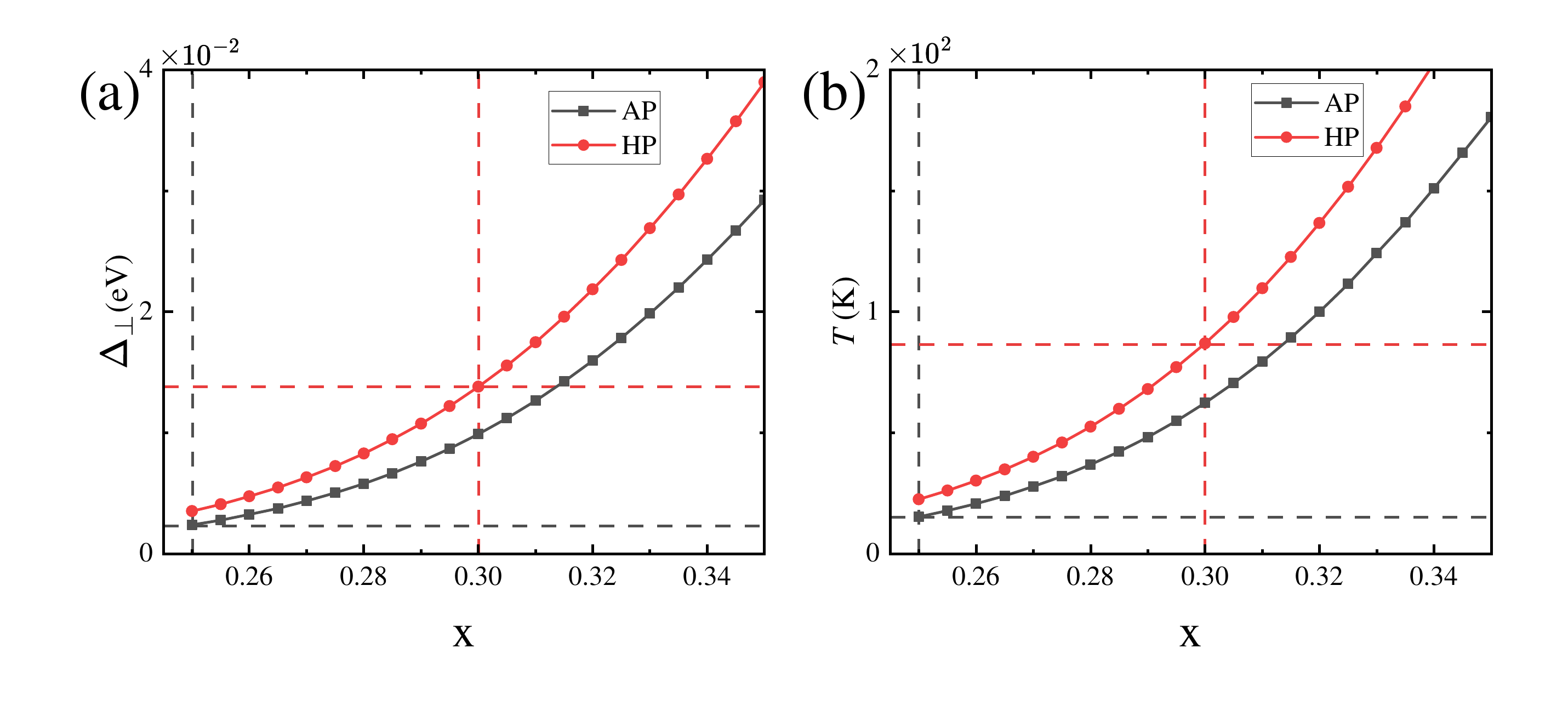}
\caption{Filling-dependent superconducting strength. (a) Inter-layer pairing gap $\Delta_\perp$ versus
filling level $x$. 
(b) Superconducting $T_c$
versus filling level $x$. The position of ambient pressure (AP) and high pressure (HP) physical filling is indicated by black dashed line ($x = 0.25$) and red dashed line ($x = 0.3$) respectively.}
\label{fig3}
\end{figure}

We apply the slave-boson mean-field (SBMF) approach \cite{kotliar1988,lee2006htsc} on the bilayer $t$-$J_{\parallel}$-$J_{\perp}$ model to investigate the SC properties under pressure. 
For a typical Hubbard interaction value of $U=4$ eV, the simulated order parameters at AP $0$ GPa and HP $30$ GPa are calculated and summarized in Table~\ref{tab:MForderParameter}. 
The filling dependence of the pairing strengths for AP ($0$ GPa) and HP ($30$ GPa) conditions is depicted in Fig.~\ref{fig3}(a).
Evidently, the inter-layer pairing amplitude $\Delta_{\perp}$ increases with increasing filling levels.
With increasing pressure, the superexchange interaction is enhanced, leading to an increasing pairing amplitude $\Delta_{\perp}$, 
from $2.40 \times 10^{-3}$ eV to $1.38 \times 10^{-2}$ eV. 
Due to the larger inter-layer superexchange, the inter-layer pairing amplitude $\Delta_{\perp}$ is significantly stronger than the intra-layer one $\Delta_{\parallel}$. 

 Our analysis reveals key characteristics of the SC state, identifying it as having a $s$-wave pairing symmetry.
For the bilayer La$_3$Ni$_2$O$_7$, the intralayer ($x$-$y$ plane) and interlayer ($z$ direction) pairing amplitudes are distinct due to the anisotropy between the $x$/$y$ and $z$ dimensions. 
There is no crystalline symmetry operation within the $D_{4h}$ point group that transform an intralayer bond into an interlayer one.
Therefore, the intralayer and interlayer pairing amplitudes together reveals an overall $A_{1g}$ pairing symmetry.
Moreover, $\Delta_{\perp}$ induces a superconducting gap that changes sign between the bonding and antibonding Fermi surfaces that arise from the bilayer crystal structure. 

Fig.~\ref{fig3}(b) presents the numerically simulated transition temperature $T_c$ for different pressures.
The enhancement of $\Delta_{\perp}$ indicates a corresponding rise in the transition temperature $T_c$. 
At the physical filling level of $0.25$, $T_c$ exhibits a dramatic increase from $15$ K at $0$ GPa to $87$ K at $30$ GPa.
Our calculations also give a ratio of $\Delta_{\perp}/T_c \approx 1.6$, consistent with the predictions of BCS theory.
Furthermore, our calculations yield a ratio of $\Delta_{\perp}/T_c \approx 1.6$, which is in consistent with the predictions by BCS theory.

Furthermore, to evaluate the impact of symmetry changes on $T_c$ within our model, we performed calculations at the physical electron filling of $x = 0.25$, eliminating the anisotropy by setting $t^{\parallel}_1$ and $t^{\parallel}_2$ to their averaged value $\sqrt{t^{\parallel}_1t^{\parallel}_2}$. 
The results show that the pairing strength $\Delta_{\perp}$ changes only slightly from $-2.40 \times 10^{-3}$ eV to $-2.59 \times 10^{-3}$ eV, and the corresponding $T_c$ increases modestly from approximately $15$ K to $16.1$ K. 
These findings indicate that symmetry changes alone are unlikely to account for the pronounced variation in superconducting $T_c$ across the structural phase transition.

\begin{figure}[t!]
\centering
\includegraphics[width=0.7\linewidth]{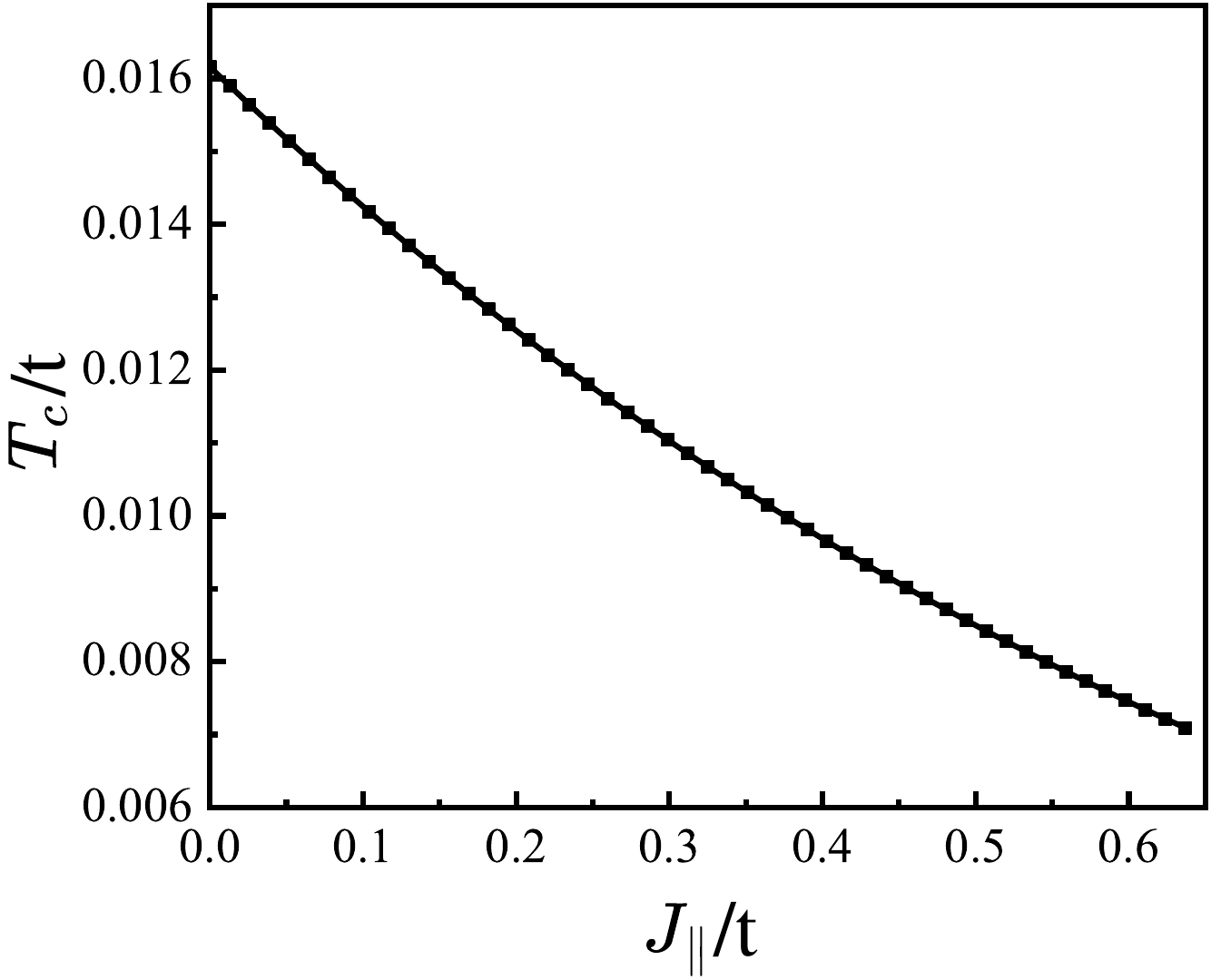}
\caption{$T_c$ versus intralayer superexchange $J_{\parallel}$, at $J_{\perp}$=0.8$t$, $x$=0.3.}
\label{fig4}
\end{figure}

Most experimental data have suggested that there is no bulk SC in the AP phase \cite{Wang2023LNO,Wang2023LNOb,YuanHQ2023LNO,wang2023LNOpoly,wang2023la2prnio7,wang2024bulk}. 
The suppression of SC at ambient pressure can be attributed to two main factors: 
First, ARPES experimental \cite{yang2024orbital} and DFT calculations \cite{YaoDX2023} both reveal that in the AP phase, the $3d_{z^2}$ orbital is exactly below the Fermi surface and will not contribute to the electronic property. 
The electron filling of the $3d_{x^2-y^2}$ orbital is well-approximated by quarter-filling, which is significantly overdoped and itinerant, reducing its electronic correlations and, consequently, the pairing strength. 
Second, at AP $Amam$ phase, the inter-layer hopping of the $3d_{z^2}$ orbital is reduced due to the deviation of the Ni-O-Ni angle $\varphi$ from 180$^{\circ}$.
This leads to a significant reduction in the inter-layer superexchange interaction $J_{\perp}$, further diminishing the pairing strength.
Experimental, robust SC of La$_3$Ni$_2$O$_7$ is found in the phase with $\varphi\approx 180^{\circ}$ \cite{bhatt2025resolving,zhong2025epitaxial}, emphasizing the important role of the inter-layer coupling.

DFT calculations ~\cite{zhang2023structural} also show that further increasing pressure decreases the ratio of the inter-layer NN hopping for the $3d_{z^2}$ orbital to the intra-layer NN hopping for the $3d_{x^2-y^2}$ orbital. 
This change directly affects the AFM superexchange interaction $J$, which scales as $4t^2/U$ in the strong-coupling limit. 
As a result, the ratio of inter-layer to intra-layer exchange couplings, $J_{\perp}/J_{\parallel}$, decreases under pressure.
Further studies indicate that the superconducting $T_c$ depends on the ratio $J_{\perp}/J_{\parallel}$: 
as $J_{\perp}/J_{\parallel}$ increases, $T_c$ rises sharply, and conversely, it drops quickly as $J_{\perp}/J_{\parallel}$ decreases ~\cite{lu2023bilayertJ}. 
This suggests that the observed reduction in $T_c$ at higher pressures can be attributed to the reduction of $J_{\perp}/J_{\parallel}$.

To further illustrate this phenomenon, we performed additional test employing the effective single-orbital $t$-$J_{\parallel}$-$J_{\perp}$ model. 
As shown in Fig.~\ref{fig4}, we fixed the hopping amplitude $t$, the inter-layer exchange coupling $J_{\perp}$, and the filling level while systematically increasing intra-layer coupling $J_{\parallel}$. 
The results show a rapid decrease in $T_c$, revealing the competition between interlayer and intralayer pairing channels. 
These findings strongly suggest that the suppression of $T_c$ under higher pressure \cite{li2024pressure} is driven by the delicate balance between interlayer and intralayer exchange couplings. 
Upon further compression, the hybridization $V$ between orbitals might become increasingly significant, potentially leading to a suppression of $T_c$ \cite{qin2023high,qu2023roles}.

\vspace{0.5\baselineskip}
\noindent{\bf Oxygen Vacancies and Superconductivity}

\begin{figure}[t!]
\centering
\includegraphics[width=0.45\textwidth]{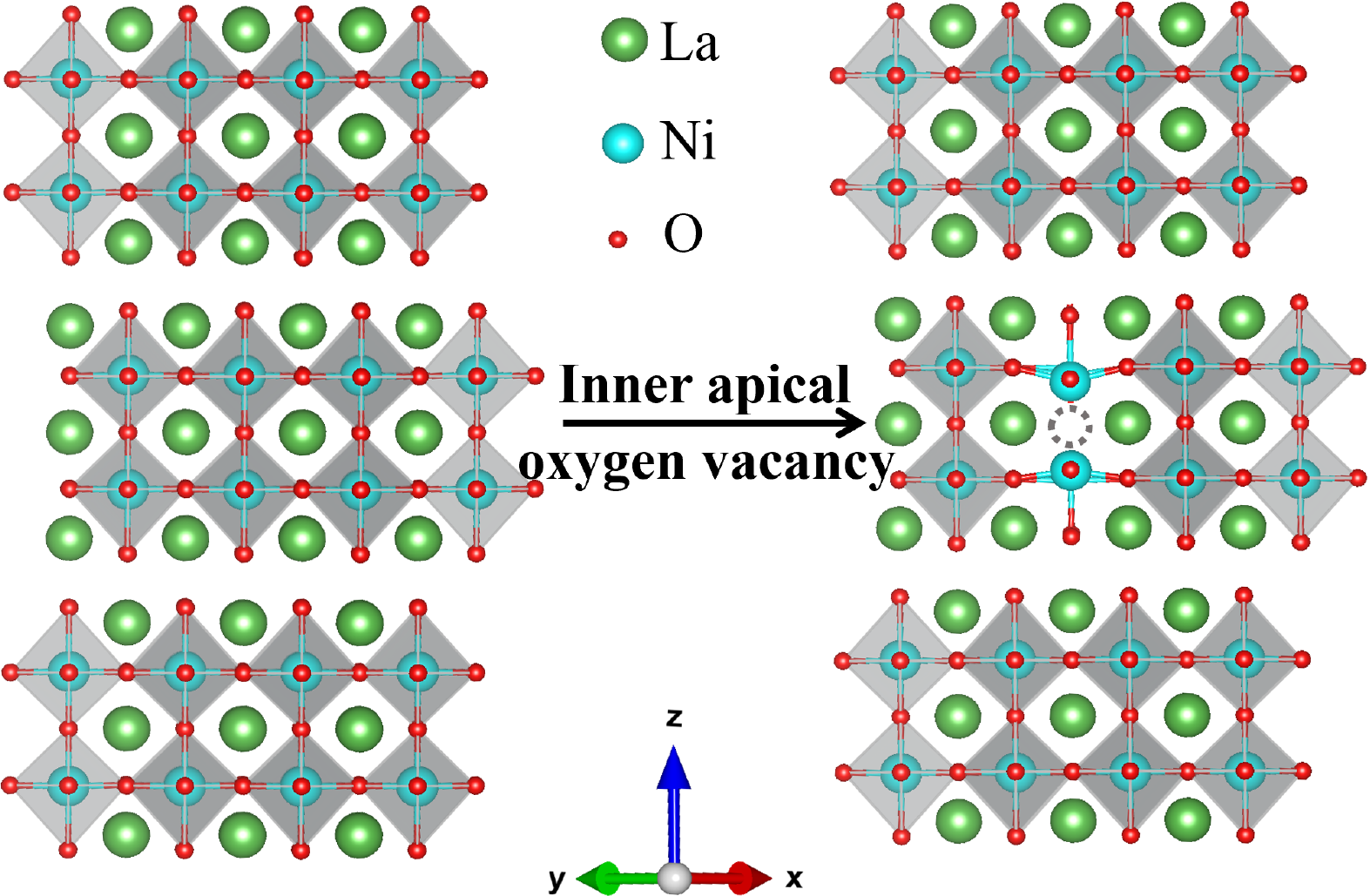}
\caption{Structural comparison between pristine La$_3$Ni$_2$O$_7$ and La$_3$Ni$_2$O$_{7-\delta}$, projected along the [110] axis. 
The left diagram displays the regular lattice structure of La$_3$Ni$_2$O$_7$, composed of alternating layers of NiO$_2$ and LaO planes. 
The right diagram illustrates La$_3$Ni$_2$O$_{7-\delta}$ with an oxygen vacancy at the inner apical position, marked by a black dashed circle. }
\label{fig5}
\end{figure}

The inter-layer apical oxygen, positioned between the two Ni-O layers, plays a critical role for the SC of La$_3$Ni$_2$O$_7$ ~\cite{Wang2023LNO,li2024pressure,ko2024signatures,liu2025superconductivity,YangF2023,zhang2023la3ni2o6}.
Experimental direct visualization studies suggest that the apical oxygen is particularly prone to vacancy formation ~\cite{Dong2024vis,li2024pressure}.
The schematic crystal structure, presented in Fig.~\ref{fig5}, illustrates the system with apical oxygen vacancies. 
The removal of apical oxygen atoms induces substantial local structural distortions, reducing the NiO$_6$ octahedra into NiO$_5$ pyramids. 
These distortions significantly alter the symmetry of the local crystal structure, resulting in anisotropic lattice constant changes: 
a marked contraction along the $c$-axis, accompanied by a slighter expansion of the in-plane lattice constants ($a$ and $b$). 
These structural modifications are expected to further influence both intralayer and interlayer electron hopping, as well as orbital hybridization.

 To investigate the effects of apical oxygen vacancies, we performed real-space SBMF calculations on a finite-size bilayer lattice. 
For a given vacancy concentration, $\delta$, a corresponding number of apical oxygen sites were chosen randomly. 
The presence of an apical oxygen vacancy disrupts the Ni-O-Ni bond path crucial for interlayer electronic processes. 
This is incorporated into our model by locally modifying the Hamiltonian: 
specifically, the interlayer hopping ($t_{\perp}$) and interlayer superexchange ($J_{\perp}$) terms associated with any Ni-Ni bond path interrupted by a vacant apical oxygen site are set to zero. 
To maintain generality and clarity, we normalize the energy scales by setting $t_{\parallel}=1$, $J_{\parallel}=0.4$ and $J_{\perp}=0.8$.

\begin{figure}[t!]
\centering
\includegraphics[width=0.4\textwidth]{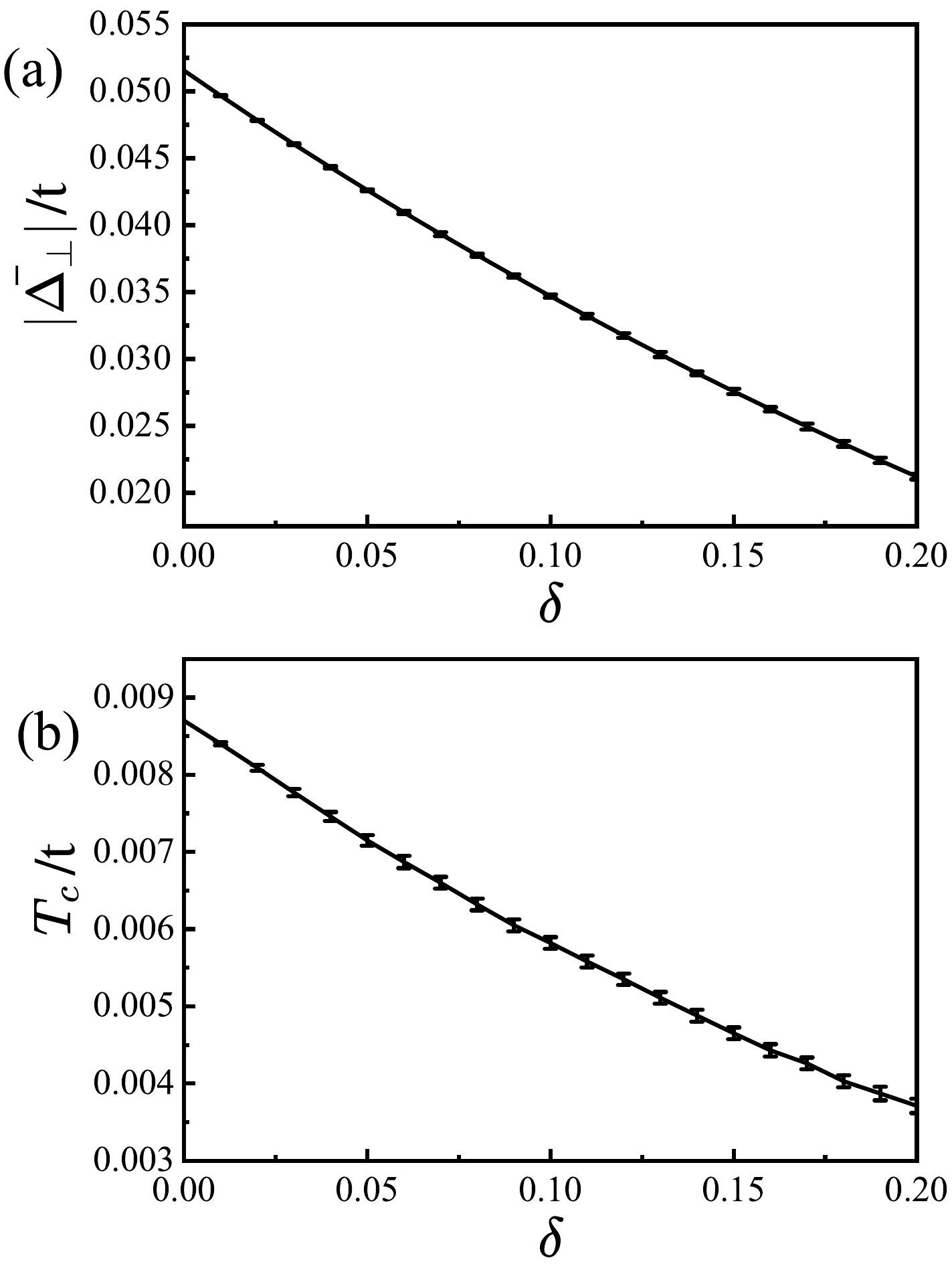}
\caption{ Dependence of pairing order parameters and superconducting $T_c$ on apical-oxygen vacancy concentrations $\delta$. (a) Dependence of pairing order parameters perpendicular to the Ni-O plane $\Delta_{\perp}$ on $\delta$. (b) Dependence of superconducting $T_c$ on $\delta$. Note that for each $\delta$, we randomly computed results for 100 samples. The solid line represents the distribution of mean values, while the error bars are the standard deviation of the deviation of each sample's data from the mean.}
\label{fig6}
\end{figure}

\begin{figure*}[t!]
\centering
\includegraphics[width=0.8\textwidth]{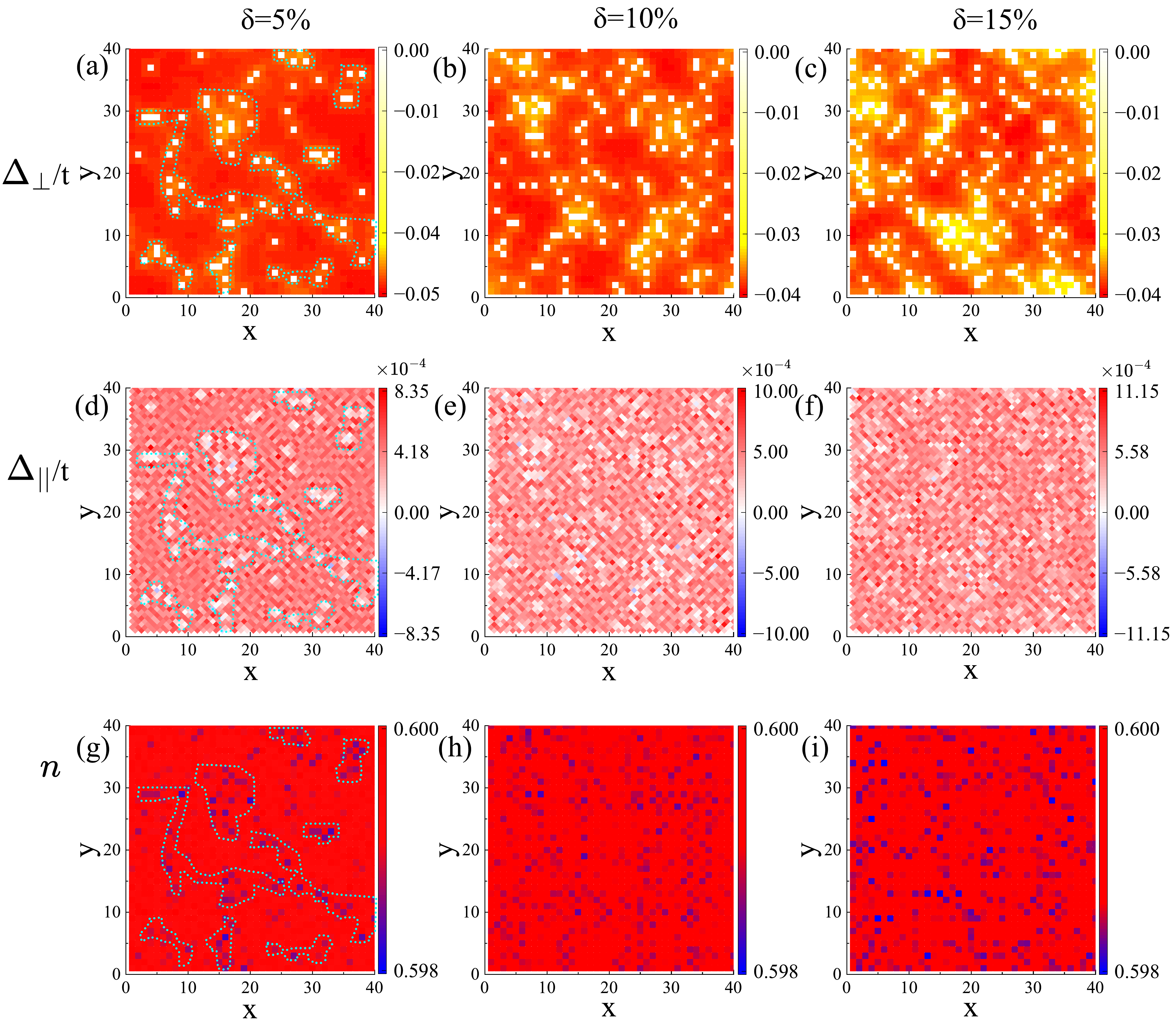}
\caption{Spatial distributions of $\Delta_{\perp}$, $\Delta_{\parallel}$, and the particle number density $n$ versus apical-oxygen vacancy $\delta$. 
The top row (a,b,c) shows the distribution of $\Delta_{\perp}$, the middle row (d,e,f) presents the distribution of $\Delta_{\parallel}$, and the bottom row (g,h,i) illustrates the spatial distribution of $n$. 
The columns correspond to different apical-oxygen vacancy concentrations: the first column (a,c,g) for $\delta=5\%$, the second column (b,e,h) for $\delta=10\%$, and the third column (c,f,i) for $\delta=15\%$. 
Color bars indicate the magnitude of each quantity, highlighting variations caused by oxygen vacancy in superconducting properties and charge distribution. 
Note that for each $\delta$, we randomly computed results for 100 samples, and the presented findings here correspond to one randomly selected configuration.}
\label{fig7}
\end{figure*}

We first compute the effects of apical oxygen vacancies on the ground state SC pairing. 
Interlayer pairing strength $\Delta_{\perp}$ remains as the dominate pairing channel.
The averaged $\Delta_{\perp}$ is depicted in Fig.~\ref{fig6}, which is significantly reduced as the concentration of apical oxygen vacancies increases. 
Fig.~\ref{fig7}(a), (b) and (c) further displays the pairing order parameters for various random configurations at oxygen vacancy concentrations of $\delta = 5\%, 10\%, 15\%$. 
It is evident that, due to the breaking of Ni–O–Ni bonds, $\Delta_{\perp}$ vanishes locally at defect sites. 
Additionally, the results highlight how $\Delta_{\perp}$ is progressively suppressed in the vicinity of these vacancies, with the suppression intensifying as $\delta$ rises. 
A similar pattern of suppression is also observed for the in-plane pairing $\Delta_{\parallel}$, as shown in Fig.~\ref{fig7} (d), (e), and (f).

The suppression of SC near oxygen vacancies is closely related to the reduction of electronic density of states (DOS). 
At the vacancy site, the disruption of the interlayer Ni-O-Ni bond reduces the electron hopping channels, thereby increasing the on-site energy of electrons at this location.
Consequently, the local electron DOS around the apical oxygen vacancy decreases, as directly illustrated in Fig.~\ref{fig7} (g), (h), and (i). 
According to BCS theory, the superconducting gap $\Delta$ and critical temperature $T_c$ scales as $\propto \exp(\frac{1}{N_FV_{eff}})$, where $N_F$ is the DOS near the Fermi level, and $V_{eff}$ is the effective interaction strength. 
Therefore, the reduction in electronic DOS near the vacancies results in a weakening of the pairing strength. 
Additionally, as detailed in the $\textbf{Supplementary}$ $\textbf{Note}$ $\textbf{1}$ Fig.~S1, the distribution of the hopping order parameter $\chi$ at various vacancy concentrations exhibits a trend similar to that of $\Delta$, further reinforcing this observation.

The dependence of the $T_c$ on the concentration of apical oxygen vacancies $\delta$, calculated using the finite-temperature SBMF framework, is presented in Fig.~\ref{fig6} (b). 
The variation in $T_c$ consistently follows the behavior of $\Delta_{\perp}$, underscoring the critical role that apical oxygen vacancies play in stabilizing the SC of La$_3$Ni$_2$O$_7$ under pressure.

\begin{figure}[t!]
\centering
\includegraphics[width=0.4\textwidth]{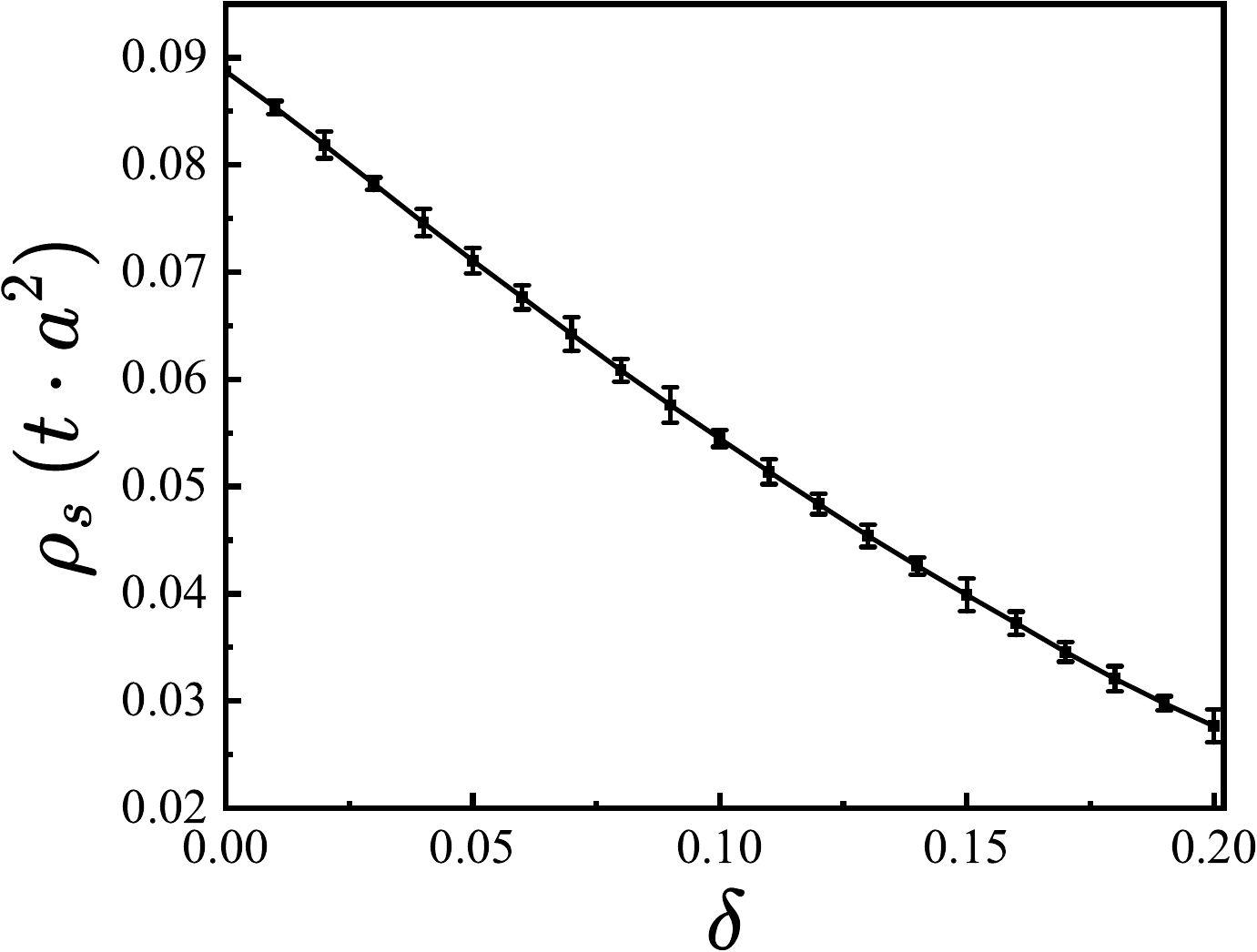}
\caption{Dependence of superfluid density $\rho_s$ on apical-oxygen vacancy concentrations $\delta$. Note that for each $\delta$, we randomly computed results for 100 samples. The solid line represents the distribution of mean values, while the error bars are the standard deviation of the deviation of each sample's data from the mean.}
\label{fig8}
\end{figure}

In addition, we explored the impact of different oxygen vacancy concentrations on the superfluid density. 
As shown in Fig.~\ref{fig8}, the superfluid density decreases rapidly with the increasing in apical oxygen vacancies. 
This phenomenon arises from two primary factors. 
On the one hand, the apical oxygen vacancies reduce the superconducting pairing strength and simultaneously decrease the diamagnetic current density. 
On the other hand, the presence of oxygen vacancies introduces randomness, which significantly enhances the paramagnetic current density. 
Consequently, the decrease in diamagnetic current density coupled with the increase in paramagnetic current density inevitably leads to a reduction in the superfluid density. 
This result is consistent with experimental measurements, which show a significantly reduced diamagnetic fraction in samples with a higher concentration of apical oxygen vacancies \cite{zhou2023investigations}.

 A reduction in superfluid density $\rho_s$ is expected to cause an increase in the magnetic penetration depth ($\lambda$), since $\rho_s\propto \lambda^{-2}$, 
consistently with weakened SC.
This would also likely manifest as a weakened Meissner response in samples with higher vacancy concentrations. 

\vspace{0.5\baselineskip}
\noindent{\bf Discussion} 
In summary, we have employed the SBMF method to investigate the effects of pressure and apical oxygen vacancies on the SC of La$_3$Ni$_2$O$_7$. 
 As pressure increases, the system undergoes a structural phase transition from the $Amam$ phase at ambient pressure to the $I_4/mmm$ phase at high pressure. 
Our calculations show a significant enhancement in the superconducting transition temperature, 
which is qualitatively consistent with experimental observations.

We further investigated the influence of apical oxygen vacancies on SC by simulating the problem within the randomness in real space. 
The introduction of apical oxygen vacancies significantly weakens SC, a suppression that is also evident in the calculated pairing order parameters and density distributions. 
These results align with experimental observations of extremely weak diamagnetic signals in many samples, particularly those with high concentrations of oxygen vacancies.

Recent experiments on bulk La$_3$Ni$_2$O$_7$ at AP show signatures of high-$T_c$ SC at $80$ K, albeit with a low volume fraction ($\sim 0.2\%$) ~\cite{huo2025low}. 
Conversely, resistivity data reveals a broad resistance drop near $20$ K, followed by a plateau at lower temperatures ~\cite{huo2025low}, hinting at a bulk-like SC phase at AP.  
Interestingly, this lower temperature is close to the theoretical predictions of weak SC around $15$ K in this work (e.g., Fig.~\ref{fig3}(b)).  
The $\sim 20$ K resistivity feature observed experimentally aligns well with our theoretical prediction of a weak, intrinsic bulk SC phase with $T_c\approx 15$K for the homogeneous Amam phase. 
Thus, the experimental data might encompass an underlying, weaker bulk SC tendency captured by our homogeneous model.

Overall, our findings provide a theoretical explanation for the experimentally observed pressure dependence of SC, including the enhancement of SC under moderate pressure and its suppression by apical oxygen vacancies. 
These insights deepen our understanding of the delicate interplay between structural factors and SC in this bilayer nickelate system.
Moreover, the perspective under AP phase points to a potential realization of AP SC in the bulk material.

 Our effective $d_{x^2-y^2}$ orbital model for La$_3$Ni$_2$O$_7$ can be considered in light of its potential charge-transfer nature and the key role of oxygen $2p$ orbitals, as suggested by experiments ~\cite{Dong2024vis}.
In this charge-transfer scenario, Ni $3d^8$ configurations can form a Zhang-Rice singlet (ZRS) \cite{zhang1988effective} by combining a Ni $E_g$ hole with a ligand $2p$ hole.
Crucially, these ZRS states differ based on the Ni orbital involved: $d_{x^2-y^2}$ orbitals form planar ZRS with O $2p_{x,y}$ orbitals (similar to cuprates), which are less affected by apical oxygens. 
In contrast, $d_{z^2}$ orbitals form out-of-plane ZRS that depend strongly on the apical oxygen $2p_z$ orbitals.

By mapping these ZRS to effective Ni $d$-holes, we can understand the impact of vacancies. 
An apical oxygen vacancy disrupts the local $d_{z^2}$-ZRS by breaking the Ni-O$_{\text{apical}}$-Ni bond, thereby significantly reducing the interlayer AFM coupling. 
This aligns directly with our model's findings, where vacancies suppress SC by weakening this critical interlayer interaction.
Therefore, the ZRS-based picture provides a justification for reducing the complex system to an effective $E_g$-orbital model, and subsequently to our $d_{x^2-y^2}$-focused approach. 
We acknowledge, however, the limitations of this effective model. A fully quantitative understanding, particularly concerning $p$-$d$ effects near vacancies, will require more comprehensive multi-orbital models including oxygen $2p$ states, which is an important direction for future research.


\vspace{0.5\baselineskip}
\noindent{\bf Methods}

\noindent{\bf Slave-boson mean-field theory}

In the slave-boson mean field theory, the $3d_{x^2-y^2}$-orbital electron operator is expressed as $c_{\bm{i}\alpha\sigma}^{\dagger}=f_{\bm{i}\alpha\sigma}^{\dagger}b_{\bm{i}\alpha}$, where $f_{\bm{i}\alpha\sigma}^{\dagger}/b_{\bm{i}\alpha}$ is the spinon/holon creation/annihilation operator, 
with the local constraint $\sum_{\sigma}f_{\bm{i}\alpha\sigma}^{\dagger} f_{\bm{i}\alpha\sigma} +b_{\bm{i}\alpha}^{\dagger} b_{\bm{i}\alpha} =1$.
Introduce the intra-layer hoppings and pairings ones,
\begin{equation}
\begin{aligned}
\chi_{\bm{i}\bm{j}}^{(\alpha)}
=f_{\bm{j}\alpha\uparrow}^{\dagger} f_{\bm{i}\alpha\uparrow}
+f_{\bm{j}\alpha\downarrow}^{\dagger} f_{\bm{i}\alpha\downarrow},\\
\Delta_{\bm{i}\bm{j}}^{(\alpha)}
=f_{\bm{j}\alpha\downarrow} f_{\bm{i}\alpha\uparrow }
-f_{\bm{j}\alpha\uparrow} f_{\bm{i}\alpha\downarrow }
\end{aligned}
\end{equation}
and the interlayer hoppings and pairings ones,
\begin{equation}
\begin{aligned}
\chi_{\bm{i},\perp}
=f_{\bm{i}2\uparrow}^{\dagger} f_{\bm{i}1\uparrow}
+f_{\bm{i}2\downarrow}^{\dagger} f_{\bm{i}1\downarrow},\\
\Delta_{\bm{i},\perp}
=f_{\bm{i}2\downarrow } f_{\bm{i}1\uparrow}
-f_{\bm{i}2\uparrow} f_{\bm{i}1\downarrow},
\end{aligned}
\end{equation}
The holon will condense at low temperature and holon operator can be replaced by its condensation density below the condensation temperature, $b_{i\alpha}\sim b_{i\alpha}^{\dagger}\sim \sqrt{\delta_{h}}=\sqrt{1-2x}$, where $x\sim 0.3$ represents the electron filling of the $d_{x^2-y^2}$-orbital under pressure-induced self-doping effects \cite{zhang2023trends}. 

The spin superexchange can be decoupled in the hopping and pairing channel, i.e., for the inter-layer one,
\begin{equation}
	\begin{aligned}
		&J_{\perp} \bm{S}_{x^2i1} \cdot \bm{S}_{x^2i2}   \\
		=&-
		\Big[\chi_{i\perp}
		\big(f_{i1\uparrow}^{\dagger} f_{i2\uparrow}
		+f_{i1\downarrow}^{\dagger} f_{i2\downarrow}\big)
		+\text{h.c.}- \frac{8|\chi_{i\perp}|^2}{3J_{\perp}}  \Big]\\ 
		&-
		\Big[\Delta_{i\perp}
		\big(f_{i1\uparrow}^{\dagger}f_{i2\downarrow}^{\dagger} 
		-f_{i1\downarrow}^{\dagger}f_{i2\uparrow}^{\dagger} \big) 
		+\text{h.c.} -\frac{8|\Delta_{i\perp}|^2}{3J_{\perp}}   \Big],
	\end{aligned}
\end{equation}
In the mean-field analysis, hoppings $\chi$, pairings $\Delta$ and Lagrange multipliers $\lambda_{\bm{i}\alpha}$ are replaced by their site-independent mean-field values,
\begin{equation}
	\begin{aligned}
		\chi_{ij}^{(\alpha)}
		=&\frac{3}{8} J_{\parallel} \langle f_{j\alpha\uparrow}^{\dagger} f_{i\alpha\uparrow}
		+f_{j\alpha\downarrow}^{\dagger} f_{i\alpha\downarrow}\rangle
		\equiv \chi_{j-i}^{(\alpha)}, \\
		\Delta_{ij}^{(\alpha)}
		=&\frac{3}{8} J_{\parallel} \langle f_{j\alpha\downarrow} f_{i\alpha\uparrow} 
		-f_{j\alpha\uparrow} f_{i\alpha\downarrow} \rangle
		\equiv \Delta_{j-i}^{(\alpha)},  \\
		\chi_{i\perp}
		=&\frac{3}{8} J_{\perp} \langle f_{i2\uparrow}^{\dagger} f_{i1\uparrow}
		+f_{i2\downarrow}^{\dagger} f_{i1\downarrow}\rangle
		\equiv \chi_{\perp}, \\
		\Delta_{i\perp}
		=&\frac{3}{8} J_{\perp} \langle f_{i2\downarrow} f_{i1\uparrow} 
		-f_{i2\uparrow} f_{i1\downarrow} \rangle
		\equiv \Delta_{\perp}.
	\end{aligned}
\end{equation}
The mean-field Hamiltonian for the spinon part is given by,
\begin{align}
&H_{f,\text{MF}}=\sum_{\bm{k}\alpha\sigma} \varepsilon_{\bm{k},\alpha} f_{\bm{k}\alpha\sigma}^{\dagger} f_{\bm{k}\alpha\sigma} \nonumber\\
&-\sum_{\bm{k}} 
\Big[\Big(\frac{3}{8} J_{\perp} \chi_{z}+t_{\perp} \delta_h \Big)  \big(f_{\bm{k}1\uparrow}^{\dagger} f_{\bm{k}2\uparrow}
+f_{-\bm{k}1\downarrow}^{\dagger} f_{-\bm{k}2\downarrow} \big)+\text{h.c.} \Big]
\nonumber \nonumber\\
&+\sum_{\bm{k}\alpha} \big(F_{\bm{k},\alpha} f_{\bm{k}\alpha\uparrow}^{\dagger} f_{-\bm{k}\alpha\downarrow}^{\dagger} +\text{h.c.} \big) \nonumber\\
&-\frac{3}{8} J_{\parallel} \sum_{\bm{k}} 
\Delta_{z} \Big[ \big(f_{\bm{k}1\uparrow}^{\dagger} f_{-\bm{k}2\downarrow}^{\dagger} 
-f_{-\bm{k}1\downarrow}^{\dagger} f_{\bm{k}2\uparrow}^{\dagger} \big) +\text{h.c.} \Big] \nonumber\\
&- \frac{3}{8} J_{\parallel} N \sum_{\alpha} \big(|\chi_{x}^{(\alpha)}|^2 +|\chi_{y}^{(\alpha)}|^2
+|\Delta_{x}^{(\alpha)}|^2 +|\Delta_{y}^{(\alpha)}|^2 \big) \nonumber\\
&+\frac{3}{8} J_{\perp} N \big(|\chi_{\perp}|^2 +|\Delta_{\perp}|^2 \big)
\end{align}
where the intralayer kinetic energy and pairing are, 
\begin{align}
\varepsilon_{\bm{k},\alpha} =& -\frac{3}{8} J_{\parallel} \big[\chi_{x}^{(\alpha)} e^{-ik_x} 
+\chi_{y}^{(\alpha)} e^{-ik_y} +\text{h.c.}\big] \nonumber\\
&-2t\delta_h \big[ \cos k_x +\cos k_y \big]  -\mu_f,    \nonumber\\
F_{\bm{k},\alpha} =& -\frac{3}{4} J_{\parallel} \big[\Delta_{x}^{(\alpha)} \cos k_x +\Delta_{y}^{(\alpha)} \cos k_y \big],
\end{align}
and $\mu_f$ is the chemical potential of the spinon field.

\vspace{0.5\baselineskip}

\noindent{\bf Data availability}

\noindent Relevant data supporting the key findings of this study are available within the article and the Supplementary Information file. 
All raw data generated during the current study are available from the corresponding authors upon request.

\vspace{0.5\baselineskip}

\noindent{\bf Code availability}

\noindent The code that supports the plots within this paper is available from the corresponding author upon request.

\bibliography{references}

@article{ko2024signatures,
  title={Signatures of ambient pressure superconductivity in thin film {L}a$_3${N}i$_2${O}$_7$},
  author={Ko, Eun Kyo and Yu, Yijun and Liu, Yidi and Bhatt, Lopa and Li, Jiarui and Thampy, Vivek and Kuo, Cheng-Tai and Wang, Bai Yang and Lee, Yonghun and Lee, Kyuho and others},
  journal={Nature},
  pages={1--2},
  year={2024},
  publisher={Nature Publishing Group UK London},
  url={https://www.nature.com/articles/s41586-024-08525-3}
}

@article{li2025photoemission,
  title={Photoemission evidence for multi-orbital hole-doping in superconducting {L}a$_{2.85}${P}r$_{0.15}${N}i$_2${O}$_7$/{S}r{L}a{A}l{O}$_4$ interfaces},
  author={Li, Peng and Zhou, Guangdi and Lv, Wei and Li, Yueying and Yue, Changming and Huang, Haoliang and Xu, Lizhi and Shen, Jianchang and Miao, Yu and Song, Wenhua and others},
  journal={arXiv:2501.09255},
  year={2025},
  url={https://arxiv.org/abs/2501.09255}
}

@article{zhou2024ambient,
  title={Ambient-pressure superconductivity onset above 40 K in bilayer nickelate ultrathin films}, 
  author={Guangdi Zhou and Wei Lv and Heng Wang and Zihao Nie and Yaqi Chen and Yueying Li and Haoliang Huang and Weiqiang Chen and Yujie Sun and Qi-Kun Xue and Zhuoyu Chen},
  year={2024},
  journal={arXiv:2412.16622},
  url={https://arxiv.org/abs/2412.16622}, 
}

@article{shi2025superconductivity,
  title={Superconductivity of the hybrid Ruddlesden-Popper {L}a$_5${N}i$_3${O}$_{11}$ single crystals under high pressure},
  author={Shi, Mengzhu and Peng, Di and Fan, Kaibao and Xing, Zhenfang and Yang, Shaohua and Wang, Yuzhu and Li, Houpu and Wu, Rongqi and Du, Mei and Ge, Binghui and others},
  journal={arXiv:2502.01018},
  year={2025},
  url={https://arxiv.org/abs/2502.01018}
}

@article{li2023trilayer,
  title={Signature of superconductivity in pressurized {L}a$_4${N}i$_3${O}$_{10}$},
  author={Li, Qing and Zhang, Ying-Jie and Xiang, Zhe-Ning and Zhang, Yuhang and Zhu, Xiyu and Wen, Hai-Hu},
  journal={Chin. Phys. Lett.},
  volume={41},
  number={1},
  pages={017401},
  year={2024},
  publisher={IOP Publishing},
  url={https://iopscience.iop.org/article/10.1088/0256-307X/41/1/017401}
}

@article{zhu2023trilayer,
  title={Superconductivity in pressurized trilayer {L}a$_4${N}i$_3${O}$_{10-\delta}$ single crystals},
  author={Zhu, Yinghao and Peng, Di and Zhang, Enkang and Pan, Bingying and Chen, Xu and Chen, Lixing and Ren, Huifen and Liu, Feiyang and Hao, Yiqing and Li, Nana and others},
  journal={Nature},
  volume={631},
  number={8021},
  pages={531--536},
  year={2024},
  publisher={Nature Publishing Group UK London},
  url={https://www.nature.com/articles/s41586-024-07553-3}
}

@article{zhang2023trilayer,
  title={Superconductivity in trilayer nickelate {L}a$_4${N}i$_3${O}$_{10}$ under pressure}, 
  author={Mingxin Zhang and Cuiying Pei and Xian Du and Yantao Cao and Qi Wang and Juefei Wu and Yidian Li and Yi Zhao and Changhua Li and Weizheng Cao and Shihao Zhu and Qing Zhang and Na Yu and Peihong Cheng and Jinkui Zhao and Yulin Chen and Hanjie Guo and Lexian Yang and Yanpeng Qi},
  journal={arXiv:2311.07423},
  url = {https://arxiv.org/abs/2311.07423},
  year={2023}
}

@article{Yuan2024la3,
  title={High-pressure crystal growth and investigation of the metal-to-metal transition of Ruddlesden–Popper trilayer nickelates {L}a$_4${N}i$_3${O}$_{10}$},
  volume={627},
  ISSN={0022-0248},
  url={http://dx.doi.org/10.1016/j.jcrysgro.2023.127511},
  DOI={10.1016/j.jcrysgro.2023.127511},
  journal={J. Cryst. Growth},
  publisher={Elsevier BV},
  author={Yuan, Ning and Elghandour, Ahmed and Arneth, Jan and Dey, Kaustav and Klingeler, Rüdiger},
  year={2024},
  month=2, 
  pages={127511} 
}

@article{li2024la3,
  title={Structural transition, electric transport, and electronic structures in the compressed trilayer nickelate {L}a$_{4}${N}i$_{3}${O}$_{10}$},
  author={Li, Jiangyuan and Chen, Cui-Qun and Huang, Chaoxin and Han, Yifeng and Huo, Mengwu and Huang, Xing and Ma, Peiyue and Qiu, Zhengyang and Chen, Junfeng and Hu, Xunwu and Chen, Lan and Xie, Tao and Shen, Bing and Sun, Hualei and Yao, Daoxin and Wang, Meng},
  journal={Sci. China- Phys. Mech. Astron.},
  year={2024},
  volume={67},
  number={11},
  pages={117403-},
  url={https://www.sciengine.com/SCPMA/doi/10.1007/s11433-023-2329-x}
}

@article{kakoi2023multiband,
author = {Kakoi, Masataka and Oi, Takashi and Ohshita, Yujiro and Yashima, Mitsuharu and Kuroki, Kazuhiko and Kato, Takeru and Takahashi, Hidefumi and Ishiwata, Shintaro and Adachi, Yoshinobu and Hatada, Naoyuki and Uda, Tetsuya and Mukuda, Hidekazu},
title = {Multiband Metallic Ground State in Multilayered Nickelates {L}a$_3${N}i$_2${O}$_7$ and {L}a$_4${N}i$_3${O}$_{10}$ Probed by $^{139}$La-NMR at Ambient Pressure},
journal = {J. Phys. Soc. Jpn.},
volume = {93},
number = {5},
pages = {053702},
year = {2024},
doi = {10.7566/JPSJ.93.053702},
URL = {https://doi.org/10.7566/JPSJ.93.053702}
}

@article{oh2024type,
  title={Type-II t-J model in charge transfer regime in bilayer La$_3$Ni$_2$O$_{7}$ and trilayer {L}a$_4${N}i$_3${O}$_{10}$},
  author={Oh, Hanbit and Zhou, Boran and Zhang, Ya-Hui},
  journal={Phys. Rev. B},
  volume={111},
  number={2},
  pages={L020504},
  year={2025},
  publisher={APS},
  url={https://journals.aps.org/prb/abstract/10.1103/PhysRevB.111.L020504}
}

@article{qin2024frustrated,
author = {Qiong Qin and Jiangfan Wang and Yi-feng Yang},
title = {Frustrated superconductivity and intrinsic reduction of $T_c$ in trilayer nickelate},
journal = {The Innovation Materials},
volume = {2},
number = {4},
pages = {100102},
year = {2024},
issn = {2959-8737},
doi = {10.59717/j.xinn-mater.2024.100102},
url = {https://www.the-innovation.org/materials/article/id/67237a399ac16a18e39f59aa}
}

@article{xu2024origin,
  title={Origin of the density wave instability in trilayer nickelate {L}a$_4${N}i$_3${O}$_{10}$ revealed by optical and ultrafast spectroscopy},
  author={Xu, Shuxiang and Chen, Cui-Qun and Huo, Mengwu and Hu, Deyuan and Wang, Hao and Wu, Qiong and Li, Rongsheng and Wu, Dong and Wang, Meng and Yao, Dao-Xin and others},
  journal={arXiv:2405.19161},
  url = {https://arxiv.org/abs/2405.19161},
  year={2024}
}

@article{du2024correlated,
  title={Correlated Electronic Structure and Density-Wave Gap in Trilayer Nickelate {L}a$_4${N}i$_3${O}$_{10}$},
  author={Du, X and Li, YD and Cao, YT and Pei, CY and Zhang, MX and Zhao, WX and Zhai, KY and Xu, RZ and Liu, ZK and Li, ZW and others},
  journal={arXiv:2405.19853},
  url = {https://arxiv.org/abs/2405.19853},
  year={2024}
}

@article{ni2024spin,
  title={Spin density wave in the bilayered nickelate La$_3$Ni$_2$O$_{7-\delta}$ at ambient pressure}, 
  author={Xiao-Sheng Ni and Yuyang Ji and Lixin He and Tao Xie and Dao-Xin Yao and Meng Wang and Kun Cao},
  year={2024},
  journal={arXiv:2407.19213},
  url={https://arxiv.org/abs/2407.19213}, 
}

@article{kakoi2023pair,
  title={Pair correlations of the hybridized orbitals in a ladder model for the bilayer nickelate {L}a$_3${N}i$_2${O}$_7$},
  author={Kakoi, Masataka and Kaneko, Tatsuya and Sakakibara, Hirofumi and Ochi, Masayuki and Kuroki, Kazuhiko},
  journal={Phys. Rev. B},
  volume={109},
  number={20},
  pages={L201124},
  year={2024},
  publisher={APS},
  url={https://journals.aps.org/prb/abstract/10.1103/PhysRevB.109.L201124}
}

@article{leonov2024la3,
  title = {Electronic structure and magnetic correlations in the trilayer nickelate superconductor {L}a$_4${N}i$_3${O}$_{10}$ under pressure},
  author = {Leonov, I. V.},
  journal = {Phys. Rev. B},
  volume = {109},
  issue = {23},
  pages = {235123},
  numpages = {7},
  year = {2024},
  month = {Jun},
  publisher = {American Physical Society},
  doi = {10.1103/PhysRevB.109.235123},
  url = {https://link.aps.org/doi/10.1103/PhysRevB.109.235123}
}

@article{tian2024effective,
doi = {10.1088/1361-648X/ad512c},
url = {https://dx.doi.org/10.1088/1361-648X/ad512c},
year = {2024},
month = {jun},
publisher = {IOP Publishing},
volume = {36},
number = {35},
pages = {355602},
author = {Tian, Peng-Fei and Ma, Hao-Tian and Ming, Xing and Zheng, Xiao-Jun and Li, Huan},
title = {Effective model and electron correlations in trilayer nickelate superconductor {L}a$_4${N}i$_3${O}$_{10}$},
journal = {J. Condens. Matter Phys.}
}

@article{wang2024nonfermi,
  title = {Non-Fermi liquid and Hund correlation in {L}a$_4${N}i$_3${O}$_{10}$ under high pressure},
  author = {Wang, Jing-Xuan and Ouyang, Zhenfeng and He, Rong-Qiang and Lu, Zhong-Yi},
  journal = {Phys. Rev. B},
  volume = {109},
  issue = {16},
  pages = {165140},
  numpages = {8},
  year = {2024},
  month = {Apr},
  publisher = {American Physical Society},
  doi = {10.1103/PhysRevB.109.165140},
  url = {https://link.aps.org/doi/10.1103/PhysRevB.109.165140}
}

@article{sakakibara2023trilayer,
  title = {Theoretical analysis on the possibility of superconductivity in the trilayer Ruddlesden-Popper nickelate {L}a$_4${N}i$_3${O}$_{10}$ under pressure and its experimental examination: Comparison with {L}a$_3${N}i$_2${O}$_7$},
  author = {Sakakibara, Hirofumi and Ochi, Masayuki and Nagata, Hibiki and Ueki, Yuta and Sakurai, Hiroya and Matsumoto, Ryo and Terashima, Kensei and Hirose, Keisuke and Ohta, Hiroto and Kato, Masaki and Takano, Yoshihiko and Kuroki, Kazuhiko},
  journal = {Phys. Rev. B},
  volume = {109},
  issue = {14},
  pages = {144511},
  numpages = {10},
  year = {2024},
  month = {Apr},
  publisher = {American Physical Society},
  doi = {10.1103/PhysRevB.109.144511},
  url = {https://link.aps.org/doi/10.1103/PhysRevB.109.144511}
}

@article{labollita2024electronic,
  title = {Electronic structure and magnetic tendencies of trilayer {L}a$_4${N}i$_3${O}$_{10}$ under pressure: Structural transition, molecular orbitals, and layer differentiation},
  author = {LaBollita, Harrison and Kapeghian, Jesse and Norman, Michael R. and Botana, Antia S.},
  journal = {Phys. Rev. B},
  volume = {109},
  issue = {19},
  pages = {195151},
  numpages = {12},
  year = {2024},
  month = {May},
  publisher = {American Physical Society},
  doi = {10.1103/PhysRevB.109.195151},
  url = {https://link.aps.org/doi/10.1103/PhysRevB.109.195151}
}

@article{yang2024effective,
  title = {Effective model and $s_\pm$-wave superconductivity in trilayer nickelate {L}a$_4${N}i$_3${O}$_{10}$},
  author = {Yang, Qing-Geng and Jiang, Kai-Yue and Wang, Da and Lu, Hong-Yan and Wang, Qiang-Hua},
  journal = {Phys. Rev. B},
  volume = {109},
  issue = {22},
  pages = {L220506},
  numpages = {7},
  year = {2024},
  month = {Jun},
  publisher = {American Physical Society},
  doi = {10.1103/PhysRevB.109.L220506},
  url = {https://link.aps.org/doi/10.1103/PhysRevB.109.L220506}
}

@article{abadi2024electronic,
  title={Electronic structure of the alternating monolayer-trilayer phase of La3Ni2O7}, 
  author={Sebastien N. Abadi and Ke-Jun Xu and Eder G. Lomeli and Pascal Puphal and Masahiko Isobe and Yong Zhong and Alexei V. Fedorov and Sung-Kwan Mo and Makoto Hashimoto and Dong-Hui Lu and Brian Moritz and Bernhard Keimer and Thomas P. Devereaux and Matthias Hepting and Zhi-Xun Shen},
  journal={arXiv:2402.07143},
  url = {https://arxiv.org/abs/2402.07143},
  year={2024}
}

@article{luo2024trilayer,
  title = {Trilayer multiorbital models of {L}a$_4${N}i$_3${O}$_{10}$},
  author = {Chen, Cui-Qun and Luo, Zhihui and Wang, Meng and W\'u, W\'ei and Yao, Dao-Xin},
  journal = {Phys. Rev. B},
  volume = {110},
  issue = {1},
  pages = {014503},
  numpages = {11},
  year = {2024},
  month = {Jul},
  publisher = {American Physical Society},
  doi = {10.1103/PhysRevB.110.014503},
  url = {https://link.aps.org/doi/10.1103/PhysRevB.110.014503}
}

@article{li2024ultrafast,
  title={Ultrafast Dynamics of Bilayer and Trilayer Nickelate Superconductors}, 
  author={Y. D. Li and Y. T. Cao and L. Y. Liu and P. Peng and H. Lin and C. Y. Pei and M. X. Zhang and H. Wu and X. Du and W. X. Zhao and K. Y. Zhai and J. K. Zhao and M. -L. Lin and P. H. Tan and Y. P. Qi and G. Li and H. J. Guo and Luyi Yang and L. X. Yang},
  journal={arXiv:2403.05012},
  url = {https://arxiv.org/abs/2403.05012},
  year={2024}
}

@article{lechermann2024electronic,
  title = {Electronic instability, layer selectivity, and Fermi arcs in {L}a$_3${N}i$_2${O}$_7$},
  author = {Lechermann, Frank and B\"otzel, Steffen and Eremin, Ilya M.},
  journal = {Phys. Rev. Mater.},
  volume = {8},
  issue = {7},
  pages = {074802},
  numpages = {7},
  year = {2024},
  month = {Jul},
  publisher = {American Physical Society},
  doi = {10.1103/PhysRevMaterials.8.074802},
  url = {https://link.aps.org/doi/10.1103/PhysRevMaterials.8.074802}
}

@article{huang2024interlayer,
  title = {Interlayer pairing-induced partially gapped Fermi surface in trilayer {L}a$_4${N}i$_3${O}$_{10}$ superconductors},
  author = {Huang, Junkang and Zhou, Tao},
  journal = {Phys. Rev. B},
  volume = {110},
  issue = {6},
  pages = {L060506},
  numpages = {6},
  year = {2024},
  month = {Aug},
  publisher = {American Physical Society},
  doi = {10.1103/PhysRevB.110.L060506},
  url = {https://link.aps.org/doi/10.1103/PhysRevB.110.L060506}
}

@article{huo2024electronic,
  title={Electronic Correlations and Hund's Rule Coupling in Trilayer Nickelate {L}a$_4${N}i$_3${O}$_{10}$},
  author={Huo, Zihao and Zhang, Peng and Zhang, Zihan and Duan, Defang and Cui, Tian},
  journal={arXiv:2407.00327},
  url = {https://arxiv.org/abs/2407.00327},
  year={2024}
}

@article{Ouyang2024absence,
author={Ouyang, Zhenfeng
and Gao, Miao
and Lu, Zhong-Yi},
title={Absence of electron-phonon coupling superconductivity in the bilayer phase of {L}a$_3${N}i$_2${O}$_7$ under pressure},
journal={npj Quantum Materials},
year={2024},
month={Oct},
day={15},
volume={9},
number={1},
pages={80},
issn={2397-4648},
doi={10.1038/s41535-024-00689-5},
url={https://doi.org/10.1038/s41535-024-00689-5}
}

@article{ryee2024quenched,
  title = {Quenched Pair Breaking by Interlayer Correlations as a Key to Superconductivity in {L}a$_{3}${N}i$_{2}${O}$_{7}$},
  author = {Ryee, Siheon and Witt, Niklas and Wehling, Tim O.},
  journal = {Phys. Rev. Lett.},
  volume = {133},
  issue = {9},
  pages = {096002},
  numpages = {7},
  year = {2024},
  month = {Aug},
  publisher = {American Physical Society},
  doi = {10.1103/PhysRevLett.133.096002},
  url = {https://link.aps.org/doi/10.1103/PhysRevLett.133.096002}
}

@article{liu2024origin,
  title={Origin of the Diagonal Double-Stripe Spin-Density-Wave and Potential Superconductivity in Bulk {L}a$_{3}${N}i$_{2}${O}$_{7}$ at Ambient Pressure},
  author={Liu, Yu-Bo and Sun, Hongyi and Zhang, Ming and Liu, Qihang and Chen, Wei-Qiang and Yang, Fan},
  journal={arXiv:2501.14752},
  year={2024},
  url={https://arxiv.org/abs/2501.14752}
}

@article{yang2024decom,
  title = {Decomposition of multilayer superconductivity with interlayer pairing},
  author = {Yang, Yi-feng},
  journal = {Phys. Rev. B},
  volume = {110},
  issue = {10},
  pages = {104507},
  numpages = {6},
  year = {2024},
  month = {Sep},
  publisher = {American Physical Society},
  doi = {10.1103/PhysRevB.110.104507},
  url = {https://link.aps.org/doi/10.1103/PhysRevB.110.104507}
}

@article{zhang2024magnetic,
  title={Magnetic Correlations and Pairing Tendencies of the Hybrid Stacking Nickelate Superlattice La$_7$Ni$_5$O$_{17}$ (La$_3$Ni$_2$O$_7$/La$_4$Ni$_3$O$_{10}$) under Pressure}, 
  author={Yang Zhang and Ling-Fang Lin and Adriana Moreo and Thomas A. Maier and Elbio Dagotto},
  journal={arXiv:2408.07690},
  year={2024},
  url={https://arxiv.org/abs/2408.07690}
}

@article{huang2024signature,
  title={Signature of Superconductivity in Pressurized Trilayer-Nickelate {P}r$_4${N}i$_3${O}$_{10-\delta}$},
  author={Huang, Xing and Zhang, Hengyuan and Li, Jingyuan and Huo, Mengwu and Chen, Junfeng and Qiu, Zhengyang and Ma, Peiyue and Huang, Chaoxin and Sun, Hualei and Wang, Meng},
  journal={Chin. Phys. Lett.},
  volume={41},
  number={12},
  pages={127403},
  year={2024},
  publisher={IOP Publishing},
  url={https://iopscience.iop.org/article/10.1088/0256-307X/41/12/127403/meta}
}

@article{liu2024evolution,
  title={Evolution of Electronic Correlations in the Ruddlesden-Popper Nickelates}, 
  author={Zhe Liu and Jie Li and Mengwu Huo and Bingke Ji and Jiahao Hao and Yaomin Dai and Mengjun Ou and Qing Li and Hualei Sun and Bing Xu and Yi Lu and Meng Wang and Hai-Hu Wen},
  journal={arXiv:2411.08539},
  year={2024},
  url={https://arxiv.org/abs/2411.08539}
}

@article{deswal2024dynamics,
  title={Dynamics of electron-electron correlated to electron-phonon coupled phase progression in trilayer nickelate La4Ni3O10}, 
  author={Sonia Deswal and Deepu Kumar and Dibyata Rout and Surjeet Singh and Pradeep Kumar},
  journal={arXiv:2411.13933},
  year={2024},
  url={https://arxiv.org/abs/2411.13933}
}

@article{zhao2024electronic,
  title={Electronic structure of Ruddlesden-Popper nickelates: strain to mimic the effects pressure}, 
  author={Yi-Feng Zhao and Antia S. Botana},
  journal={arXiv:2412.04391},
  year={2024},
  url={https://arxiv.org/abs/2412.04391}
}

@article{taniguchi1995transport,
  title={Transport, magnetic and thermal properties of {L}a$_3${N}i$_2${O}$_{7-\delta}$},
  author={Taniguchi, Satoshi and Nishikawa, Takashi and Yasui, Yukio and Kobayashi, Yoshiaki and Takeda, Jun and Shamoto, Shin-ichi and Sato, Masatoshi},
  journal={J. Phys. Soc. Jpn.},
  volume={64},
  number={5},
  pages={1644--1650},
  year={1995},
  url={https://journals.jps.jp/doi/10.1143/JPSJ.64.1644},
  publisher={The Physical Society of Japan},
}

@article{seo1996electronic,
  title={Electronic band structure and Madelung potential study of the Nickelates {L}a$_2${N}i{O}$_{4}$, {L}a$_3${N}i$_2${O}$_{7}$, and {L}a$_4${N}i$_3${O}$_{10}$},
  author={Seo, D-K and Liang, W and Whangbo, M-H and Zhang, Z and Greenblatt, M},
  journal={Inorg. Chem.},
  volume={35},
  number={22},
  pages={6396--6400},
  year={1996},
  url={https://pubs.acs.org/doi/full/10.1021/ic960379j},
  publisher={ACS Publications}
}

@article{kobayashi1996transport,
  title={Transport and magnetic properties of {L}a$_3${N}i$_2${O}$_{7-\delta}$ and {L}a$_4${N}i$_3${O}$_{10-\delta}$},
  author={Kobayashi, Yoshiaki and Taniguchi, Satoshi and Kasai, Mayumi and Sato, Masatoshi and Nishioka, Takashi and Kontani, Masaaki},
  journal={J. Phys. Soc. Jpn.},
  volume={65},
  number={12},
  pages={3978--3982},
  year={1996},
  url={https://journals.jps.jp/doi/10.1143/JPSJ.65.3978},
  publisher={The Physical Society of Japan}
}

@article{greenblatt1997ruddlesden,
  title={Ruddlesden-Popper {L}n$_{n+1}${N}i$_n${O}$_{3n+1}$ nickelates: structure and properties},
  author={Greenblatt, Martha},
  journal={Curr. Opin. Solid State Mater. Sci.},
  volume={2},
  number={2},
  pages={174--183},
  year={1997},
  doi = {https://doi.org/10.1016/S1359-0286(97)80062-9},
  url = {https://www.sciencedirect.com/science/article/pii/S1359028697800629},
  publisher={Elsevier}
}

@article{greenblatt1997electronic,
  title={Electronic properties of {L}a$_3${N}i$_2${O}$_7$ and {L}n$_4${N}i$_3${O}$_{10}$, {L}n= {L}a, {P}r and {N}d},
  author={Greenblatt, M and Zhang, Z and Whangbo, MH},
  journal={Synth. Met.},
  volume={85},
  number={1-3},
  pages={1451--1452},
  year={1997},
  doi = {https://doi.org/10.1016/S0379-6779(97)80312-8},
  url = {https://www.sciencedirect.com/science/article/pii/S0379677997803128},
  publisher={Elsevier}
}

@article{ling2000neutron,
  title={Neutron diffraction study of {L}a$_3${N}i$_2${O}$_7$: Structural relationships among $n=$1,2, and 3 phases {L}a$_{n+1}${N}i$_n${O}$_{3n+1}$},
  author={Ling, Christopher D and Argyriou, Dimitri N and Wu, Guoqing and Neumeier, JJ},
  journal={J. Solid State Chem.},
  volume={152},
  number={2},
  pages={517--525},
  year={2000},
  doi={https://doi.org/10.1006/jssc.2000.8721},
  url={https://www.sciencedirect.com/science/article/pii/S0022459600987218},
  publisher={Elsevier}
}

@article{wu2001magnetic,
  title={Magnetic susceptibility, heat capacity, and pressure dependence of the electrical resistivity of {L}a$_3${N}i$_2${O}$_7$ and {L}a$_4${N}i$_3${O}$_{10}$},
  author={Wu, Guoqing and Neumeier, J. J and Hundley, M. F.},
  journal={Phys. Rev. B},
  volume={63},
  number={24},
  pages={245120},
  year={2001},
  url={https://journals.aps.org/prb/abstract/10.1103/PhysRevB.63.245120},
  publisher={APS}
}

@article{fukamachi2001nmr,
  title={$^{139}${L}a {NMR} studies of layered perovskite systems {L}a$_3${N}i$_2${O}$_{7-\delta}$ and {L}a$_4${N}i$_3${O}$_{10}$},
  author={Fukamachi, T and Kobayashi, Y and Miyashita, T and Sato, M},
  journal={J. Phys. Chem. Solids},
  volume={62},
  number={1-2},
  pages={195--198},
  year={2001},
  url={https://www.sciencedirect.com/science/article/pii/S002236970000127X},
  publisher={Elsevier}
}

@article{voronin2001neutron,
  title={Neutron diffraction, synchrotron radiation and EXAFS spectroscopy study of crystal structure peculiarities of the lanthanum nickelates {L}a$_{n+1}${N}i$_n${O}$_y$ ($n=1,2,3$)},
  author={Voronin, V. I. and Berger, I. F. and Cherepanov, V. A. and Gavrilova, L. Ya. and Petrov, A. N. and Ancharov, A. I. and Tolochko, B.P. and Nikitenko, S. G.},
  journal={Nucl. Instrum. Methods Phys. Res. A},
  volume={470},
  number={1-2},
  pages={202--209},
  year={2001},
  url={https://www.sciencedirect.com/science/article/pii/S0168900201010361},
  publisher={Elsevier}
}

@article{xie2024strong,
  title={Strong interlayer magnetic exchange coupling in La3Ni2O7- $\delta$ revealed by inelastic neutron scattering},
  author={Xie, Tao and Huo, Mengwu and Ni, Xiaosheng and Shen, Feiran and Huang, Xing and Sun, Hualei and Walker, Helen C and Adroja, Devashibhai and Yu, Dehong and Shen, Bing and others},
  journal={Science Bulletin},
  volume={69},
  number={20},
  pages={3221--3227},
  year={2024},
  publisher={Elsevier},
  url={https://www.sciencedirect.com/science/article/pii/S2095927324005164}
}

@article{Bannikov2006,
author = {D. O. Bannikov and A. P. Safronov and V. A. Cherepanov},
title = {Thermochemical characteristics of {L}a$_{n+1}${N}i$_n${O}$_{3n+1}$ oxides},
journal = {Thermochim. Acta},
volume = {451},
number = {1},
pages = {22-26},
year = {2006},
issn = {0040-6031},
doi = {https://doi.org/10.1016/j.tca.2006.08.004},
url = {https://www.sciencedirect.com/science/article/pii/S0040603106004424},
}

@article{hosoya2008pressure,
title = {Pressure studies on the electrical properties in {R}$_{2-x}${S}r$_x${N}i$_{1-y}${C}u$_y${O}$_{4+\delta}$ ({R}={L}a, {N}d) and {L}a$_3${N}i$_2${O}$_{7+\delta}$},
author={Hosoya, T and Igawa, K and Takeuchi, Y and Yoshida, K and Uryu, T and Hirabayashi, H and Takahashi, H},
doi = {10.1088/1742-6596/121/5/052013},
url={https://iopscience.iop.org/article/10.1088/1742-6596/121/5/052013/meta},
year = {2008},
month = {jul},
publisher = {},
volume = {121},
number = {5},
pages = {052013},
journal = {J. Phys.: Conf. Ser.},
}

@article{pardo2011dft,
  title={Metal-insulator transition in layered nickelates {L}a$_3${N}i$_2${O}$_{7-\delta}$ ($\delta$= 0.0, 0.5, 1)},
  author={Pardo, Victor and Pickett, Warren E},
  journal = {Phys. Rev. B},
  volume={83},
  number={24},
  pages={245128},
  year={2011},
  publisher = {American Physical Society},
  doi = {10.1103/PhysRevB.83.245128},
  url = {https://link.aps.org/doi/10.1103/PhysRevB.83.245128}
}

@article{nakata2017finite,
  title={Finite-energy spin fluctuations as a pairing glue in systems with coexisting electron and hole bands},
  author={Nakata, Masahiro and Ogura, Daisuke and Usui, Hidetomo and Kuroki, Kazuhiko},
  journal={Phys. Rev. B},
  volume={95},
  number={21},
  pages={214509},
  year={2017},
  publisher={APS},
  url = {https://link.aps.org/doi/10.1103/PhysRevB.95.214509}
}

@article{mochizuki2018strain,
  title={Strain-engineered Peierls instability in layered perovskite {L}a$_3${N}i$_2${O}$_7$ from first principles},
  author={Mochizuki, Yasuhide and Akamatsu, Hirofumi and Kumagai, Yu and Oba, Fumiyasu},
  journal={Phys. Rev. Mater.},
  volume={2},
  number={12},
  pages={125001},
  year={2018},
  url={https://journals.aps.org/prmaterials/abstract/10.1103/PhysRevMaterials.2.125001},
  publisher={APS}
}

@article{li2020epitaxial,
  title={Epitaxial growth and electronic structure of Ruddlesden--Popper nickelates ({L}a$_{n+1}${N}i$_n${O}$_{3n+1}$, $n=$ 1--5)},
  author={Li, Z and Guo, W and Zhang, T. T. and Song, J. H. and Gao, T. Y. and Gu, Z. B. and Nie, Y. F.},
  journal={APL Mater.},
  volume = {8},
  number = {9},
  pages = {091112},
  year={2020},
  url={https://pubs.aip.org/aip/apm/article/8/9/091112/122963},
  publisher={AIP Publishing}
}

@article{song2020structure,
  title={Structure, electrical conductivity and oxygen transport properties of Ruddlesden--Popper phases {L}n$_{n+1}${N}i$_n${O}$_{3n+1}$ ({L}n={L}a, {P}r and {N}d; $n$=1, 2 and 3)},
  author={Song, Jia and Ning, De and Boukamp, Bernard and Bassat, Jean-Marc and Bouwmeester, Henny JM},
  journal={J. Mater. Chem. A},
  volume={8},
  number={42},
  pages={22206--22221},
  year={2020},
  url={https://pubs.rsc.org/en/content/articlelanding/2020/ta/d0ta06731h},
  publisher={Royal Society of Chemistry}
}

@article{barone2021improved,
  title={Improved control of atomic layering in perovskite-related homologous series},
  author = {Barone, Matthew R. and Dawley, Natalie M. and Nair, Hari P. and Goodge, Berit H. and Holtz, Megan E. and Soukiassian, Arsen and Fleck, Erin E. and Lee, Kiyoung and Jia, Yunfa and Heeg, Tassilo and Gatt, Refael and Nie, Yuefeng and Muller, David A. and Kourkoutis, Lena F. and Schlom, Darrell G.},
  journal={APL Mater.},
  volume={9},
  number={2},
  pages = {021118},
  year={2021},
  url={https://pubs.aip.org/aip/apm/article/9/2/021118/567757/Improved-control-of-atomic-layering-in-perovskite},
  publisher={AIP Publishing}
}

@article{Wang2022LNO,
  title={Evidence for charge and spin density waves in single crystals of {L}a$_3${N}i$_2${O}$_7$ and {L}a$_3${N}i$_2${O}$_6$},
  author={Liu, Zengjia and Sun, Hualei and Huo, Mengwu and Ma, Xiaoyan and Ji, Yi and Yi, Enkui and Li, Lisi and Liu, Hui and Yu, Jia and Zhang, Ziyou and Chen, Zhiqiang and Liang, Feixiang and Dong, Hongliang and Guo, Hanjie and Zhong, Dingyong and Shen, Bing and Li, Shiliang and Wang, Meng},
  journal={Sci. China-Phys. Mech. Astron.},
  volume={66},
  number={1},
  pages={217411},
  year={2023},
  publisher={Springer},
  url = {https://link.springer.com/article/10.1007/s11433-022-1962-4}
}

@article{Wang2023LNO,
   author = {Sun, Hualei and Huo, Mengwu and Hu, Xunwu and Li, Jingyuan and Liu, Zengjia and Han, Yifeng and Tang, Lingyun and Mao, Zhongquan and Yang, Pengtao and Wang, Bosen and Cheng, Jinguang and Yao, Dao-Xin and Zhang, Guang-Ming and Wang, Meng},
   title = {Signatures of superconductivity near $80${K} in a nickelate under high pressure},
journal={Nature},
year={2023},
month={Sep},
day={01},
volume={621},
number={7979},
pages={493-498},
issn={1476-4687},
doi={10.1038/s41586-023-06408-7},
url={https://doi.org/10.1038/s41586-023-06408-7}
}

@article{YuanHQ2023LNO,
author={Zhang, Yanan
and Su, Dajun
and Huang, Yanen
and Shan, Zhaoyang
and Sun, Hualei
and Huo, Mengwu
and Ye, Kaixin
and Zhang, Jiawen
and Yang, Zihan
and Xu, Yongkang
and Su, Yi
and Li, Rui
and Smidman, Michael
and Wang, Meng
and Jiao, Lin
and Yuan, Huiqiu},
title={High-temperature superconductivity with zero resistance and strange-metal behaviour in {L}a$_3${N}i$_2${O}$_{7-\delta}$},
journal={Nat. Phys.},
year={2024},
month={Aug},
day={01},
volume={20},
number={8},
pages={1269-1273},
issn={1745-2481},
doi={10.1038/s41567-024-02515-y},
}

@article{liu2024electronic,
author={Liu, Zhe and Huo, Mengwu and Li, Jie and Li, Qing and Liu, Yuecong and Dai, Yaomin and Zhou, Xiaoxiang and Hao, Jiahao and Lu, Yi and Wang, Meng and Wen, Hai-Hu},
title={Electronic correlations and partial gap in the bilayer nickelate {L}a$_3${N}i$_2${O}$_7$},
journal={Nat. Commun.},
year={2024},
month={Aug},
day={31},
volume={15},
number={1},
pages={7570},
issn={2041-1723},
doi={10.1038/s41467-024-52001-5},
url={https://www.nature.com/articles/s41467-024-52001-5}
}

@article{Dong2024vis,
author={Dong, Zehao
and Huo, Mengwu
and Li, Jie
and Li, Jingyuan
and Li, Pengcheng
and Sun, Hualei
and Gu, Lin
and Lu, Yi
and Wang, Meng
and Wang, Yayu
and Chen, Zhen},
title={Visualization of oxygen vacancies and self-doped ligand holes in {L}a$_3${N}i$_2${O}$_{7-\delta}$},
journal={Nature},
year={2024},
month={Jun},
day={01},
volume={630},
number={8018},
pages={847-852},
issn={1476-4687},
doi={10.1038/s41586-024-07482-1},
url={https://doi.org/10.1038/s41586-024-07482-1}
}

@article{wang2024bulk,
title={Bulk high-temperature superconductivity in the high-pressure tetragonal phase of bilayer {L}a$_2${P}r{N}i$_2${O}$_7$}, 
author={Ningning Wang and Gang Wang and Xiaoling Shen and Jun Hou and Jun Luo and Xiaoping Ma and Huaixin Yang and Lifen Shi and Jie Dou and Jie Feng and Jie Yang and Yunqing Shi and Zhian Ren and Hanming Ma and Pengtao Yang and Ziyi Liu and Yue Liu and Hua Zhang and Xiaoli Dong and Yuxin Wang and Kun Jiang and Jiangping Hu and Stuart Calder and Jiaqiang Yan and Jianping Sun and Bosen Wang and Rui Zhou and Yoshiya Uwatoko and Jinguang Cheng},
journal={Nature},
year={2024},
month={Oct},
day={01},
volume={634},
number={8034},
pages={579-584},
issn={1476-4687},
doi={10.1038/s41586-024-07996-8},
url={https://doi.org/10.1038/s41586-024-07996-8}
}

@article{li2024pressure,
title={Pressure-driven right-triangle shape superconductivity in bilayer nickelate {L}a$_3${N}i$_2${O}$_7$}, 
author={Jingyuan Li and Peiyue Ma and Hengyuan Zhang and Xing Huang and Chaoxin Huang and Mengwu Huo and Deyuan Hu and Zixian Dong and Chengliang He and Jiahui Liao and Xiang Chen and Tao Xie and Hualei Sun and Meng Wang},
year={2024},
journal={arXiv:2404.11369},
url={https://arxiv.org/abs/2404.11369}, 
}

@article{wang2023LNOpoly,
  title = {Pressure-Induced Superconductivity In Polycrystalline {L}a$_3${N}i$_2${O}$_7$},
  author = {Wang, G. and Wang, N. N. and Shen, X. L. and Hou, J. and Ma, L. and Shi, L. F. and Ren, Z. A. and Gu, Y. D. and Ma, H. M. and Yang, P. T. and Liu, Z. Y. and Guo, H. Z. and Sun, J. P. and Zhang, G. M. and Calder, S. and Yan, J.-Q. and Wang, B. S. and Uwatoko, Y. and Cheng, J.-G.},
  journal = {Phys. Rev. X},
  volume = {14},
  issue = {1},
  pages = {011040},
  numpages = {8},
  year = {2024},
  month = {Mar},
  publisher = {American Physical Society},
  doi = {10.1103/PhysRevX.14.011040},
  url = {https://link.aps.org/doi/10.1103/PhysRevX.14.011040}
}

@article{liu2025superconductivity,
  title={Superconductivity and normal-state transport in compressively strained {L}a$_2${P}r{N}i$_2${O}$_7$ thin films},
  author={Liu, Yidi and Ko, Eun Kyo and Tarn, Yaoju and Bhatt, Lopa and Goodge, Berit H and Muller, David A and Raghu, Srinivas and Yu, Yijun and Hwang, Harold Y},
  journal={arXiv:2501.08022},
  year={2025},
  url={https://arxiv.org/abs/2501.08022}
}

@article{Wang2023LNOb,
   author = {Jun Hou and Peng-Tao Yang and Zi-Yi Liu and Jing-Yuan Li and Peng-Fei Shan and Liang Ma and Gang Wang and Ning-Ning Wang and Hai-Zhong Guo and Jian-Ping Sun and Yoshiya Uwatoko and Meng Wang and Guang-Ming Zhang and Bo-Sen Wang and Jin-Guang Cheng},
   title = {Emergence of High-Temperature Superconducting Phase in Pressurized {L}a$_{3}${N}i$_{2}${O}$_7$ Crystals},
   publisher = {Chin. Phys. Lett.},
   year = {2023},
   journal = {Chin. Phys. Lett.},
   volume = {40},
   number = {11},
   eid = {117302},
   pages = {117302},
   url = {https://cpl.iphy.ac.cn/EN/abstract/article_116425.shtml},
   doi = {10.1088/0256-307X/40/11/117302}
}

@article{yang2024orbital,
  title={Orbital-dependent electron correlation in double-layer nickelate {L}a$_3${N}i$_2${O}$_7$},
  author={Yang, Jiangang and Sun, Hualei and Hu, Xunwu and Xie, Yuyang and Miao, Taimin and Luo, Hailan and Chen, Hao and Liang, Bo and Zhu, Wenpei and Qu, Gexing and Cui-Qun Chen and Mengwu Huo and Yaobo Huang and Shenjin Zhang and Fengfeng Zhang and Feng Yang and Zhimin Wang and Qinjun Peng and Hanqing Mao and Guodong Liu and Zuyan Xu and Tian Qian and Dao-Xin Yao and Meng Wang and Lin Zhao and X. J. Zhou},
  journal={Nat. Commun.},
  volume={15},
  number={1},
  pages={4373},
  year={2024},
  publisher={Nature Publishing Group UK London},
  url={https://www.nature.com/articles/s41467-024-48701-7}
}

@article{dan2024spin,
  title={Spin-density-wave transition in double-layer nickelate {L}a$_3${N}i$_2${O}$_7$},
  author={Dan, Zhao and Zhou, Yanbing and Huo, Mengwu and Wang, Yu and Nie, Linpeng and Wang, Meng and Wu, Tao and Chen, Xianhui},
  journal={arXiv:2402.03952},
  year={2024},
  url={https://arxiv.org/abs/2402.03952}
}

@article{chen2024electronic,
  title={Electronic and magnetic excitations in La3Ni2O7},
  author={Chen, Xiaoyang and Choi, Jaewon and Jiang, Zhicheng and Mei, Jiong and Jiang, Kun and Li, Jie and Agrestini, Stefano and Garcia-Fernandez, Mirian and Sun, Hualei and Huang, Xing and others},
  journal={Nature communications},
  volume={15},
  number={1},
  pages={9597},
  year={2024},
  publisher={Nature Publishing Group UK London},
  url={https://www.nature.com/articles/s41467-024-53863-5}
}

@article{zhou2023investigations,
  title={Investigations of key issues on the reproducibility of high-$T_c$ superconductivity emerging from compressed {L}a$_3${N}i$_2${O}$_7$},
  author={Zhou, Yazhou and Guo, Jing and Cai, Shu and Sun, Hualei and Li, Chengyu and Zhao, Jinyu and Wang, Pengyu and Han, Jinyu and Chen, Xintian and Chen, Yongjin and others},
  journal={Matter and Radiation at Extremes},
  volume={10},
  number={2},
  year={2025},
  publisher={AIP Publishing},
  url={https://pubs.aip.org/aip/mre/article/10/2/027801/3331819}
}

@article{wang2023structure,
  title={Structure Responsible for the Superconducting State in {L}a$_3${N}i$_2${O}$_7$ at High-Pressure and Low-Temperature Conditions},
  author={Wang, Luhong and Li, Yan and Xie, Sheng-Yi and Liu, Fuyang and Sun, Hualei and Huang, Chaoxin and Gao, Yang and Nakagawa, Takeshi and Fu, Boyang and Dong, Bo and others},
  journal={J. Am. Chem. Soc.},
  volume={146},
  number={11},
  pages={7506--7514},
  year={2024},
  publisher={ACS Publications},
  url={https://pubs.acs.org/doi/abs/10.1021/jacs.3c13094}
}

@article{zhang2023pressure,
  title={Effects of pressure and doping on Ruddlesden-Popper phases {L}a$_{n+1}${N}i$_n${O}$_{3n+1}$},
  author={Zhang, Mingxin and Pei, Cuiying and Wang, Qi and Zhao, Yi and Li, Changhua and Cao, Weizheng and Zhu, Shihao and Wu, Juefei and Qi, Yanpeng},
  journal={J. Mater. Res. Technol.},
  volume={185},
  pages={147--154},
  year={2024},
  publisher={Elsevier},
  url={https://www.sciencedirect.com/science/article/pii/S1005030223009829}
}

@article{wang2023la2prnio7,
  title={Observation of high-temperature superconductivity in the high-pressure tetragonal phase of {L}a$_2${P}r{N}i$_2${O}$_{7-\delta}$}, 
  author={Gang Wang and Ningning Wang and Yuxin Wang and Lifen Shi and Xiaoling Shen and Jun Hou and Hanming Ma and Pengtao Yang and Ziyi Liu and Hua Zhang and Xiaoli Dong and Jianping Sun and Bosen Wang and Kun Jiang and Jiangping Hu and Yoshiya Uwatoko and Jinguang Cheng},
  journal={arXiv:2311.08212},
  url = {https://arxiv.org/abs/2311.08212},
  year={2023}
}

@article{cui2024strain,
  title={Strain-mediated phase crossover in Ruddlesden--Popper nickelates},
  author={Cui, Ting and Choi, Songhee and Lin, Ting and Liu, Chen and Wang, Gang and Wang, Ningning and Chen, Shengru and Hong, Haitao and Rong, Dongke and Wang, Qianying and others},
  journal={Commun. Mater.},
  volume={5},
  number={1},
  pages={32},
  year={2024},
  publisher={Nature Publishing Group UK London},
  url={https://www.nature.com/articles/s43246-024-00478-4}
}

@article{li2024electronic,
  title={Electronic correlation and pseudogap-like behavior of high-temperature superconductor {L}a$_3${N}i$_2${O}$_7$},
  author={Li, Yidian and Du, Xian and Cao, Yantao and Pei, Cuiying and Zhang, Mingxin and Zhao, Wenxuan and Zhai, Kaiyi and Xu, Runzhe and Liu, Zhongkai and Li, Zhiwei and others},
  journal={Chin. Phys. Lett.},
  volume={41},
  number={8},
  pages={087402},
  year={2024},
  publisher={IOP Publishing},
  url={https://iopscience.iop.org/article/10.1088/0256-307X/41/8/087402/}
}

@article{zhang2024doping,
  title={Doping evolution of the normal state magnetic excitations in pressurized {L}a$_3${N}i$_2${O}$_7$},
  author={Zhang, Hai-Yang and Bai, Yu-Jie and Kong, Fan-Jie and Wu, Xiu-Qiang and Xing, Yu-Heng and Xu, Ning},
  journal={New J. Phys.},
  volume={26},
  number={12},
  pages={123027},
  year={2024},
  publisher={IOP Publishing},
  url={https://iopscience.iop.org/article/10.1088/1367-2630/ada0d4/meta}
}

@article{ren2025resolving,
  title={Resolving the electronic ground state of {L}a$_3${N}i$_2${O}$_{7-\delta}$ films},
  author={Ren, Xiaolin and Sutarto, Ronny and Wu, Xianxin and Zhang, Jianfeng and Huang, Hai and Xiang, Tao and Hu, Jiangping and Comin, Riccardo and Zhou, Xingjiang and Zhu, Zhihai},
  journal={Commun. Phys.},
  volume={8},
  number={1},
  pages={52},
  year={2025},
  publisher={Nature Publishing Group UK London},
  url={https://www.nature.com/articles/s42005-025-01971-z}
}

@article{li2024distinguishing,
  title={Distinguishing Electronic Band Structure of Single-layer and Bilayer Ruddlesden-Popper Nickelates Probed by in-situ High Pressure X-ray Absorption Near-edge Spectroscopy},
  author={Li, Mingtao and Wang, Yiming and Pei, Cuiying and Zhang, Mingxin and Li, Nana and Guan, Jiayi and Amboage, Monica and Adama, N and Kong, Qingyu and Qi, Yanpeng and others},
  journal={arXiv:2410.04230},
  year={2024},
  url={https://arxiv.org/abs/2410.04230}
}

@article{zhou2024revealing,
  title={Revealing nanoscale structural phase separation in {L}a$_3${N}i$_2${O}$_{7-\delta}$ single crystal via scanning near-field optical microscopy},
  author={Zhou, Xiaoxiang and He, Weihong and Zhou, Zijian and Ni, Kaipeng and Huo, Mengwu and Hu, Deyuan and Zhu, Yinghao and Zhang, Enkang and Jiang, Zhicheng and Zhang, Shuaikang and others},
  journal={arXiv:2410.06602},
  year={2024},
  url={https://arxiv.org/abs/2410.06602}
}

@article{su2024strongly,
  title={Strongly Anisotropic Charge Dynamics in La3Ni2O7 with Coherent-to-Incoherent Crossover of Interlayer Charge Dynamics},
  author={Su, Bo and Huang, Chaoxin and Zhao, Jianzhou and Huo, Mengwu and Luo, Jianlin and Wang, Meng and Chen, Zhi-Guo},
  journal={arXiv:2411.10786},
  year={2024},
  url={https://arxiv.org/abs/2411.10786}
}

@article{mijit2024local,
  title={Local electronic properties of {L}a$_3${N}i$_2${O}$_{7}$ under pressure},
  author={Mijit, Emin and Ma, Peiyue and Sahle, Christoph J and Rosa, Angelika D and Hu, Zhiwei and De Angelis, Francesco and Lopez, Alberto and Amatori, Simone and Tchoudinov, Georghii and Joly, Yves and others},
  journal={arXiv:2412.08269},
  year={2024},
  url={https://arxiv.org/abs/2412.08269}
}

@article{chen2024unveiling,
  title={Unveiling the multiband metallic nature of the normal state in nickelate {L}a$_3${N}i$_2${O}$_{7}$},
  author={Chen, Bowen and Zhang, Hengyuan and Li, Jingyuan and Hu, Deyuan and Huo, Mengwu and Wang, Shuyang and Xi, Chuanying and Wang, Zhaosheng and Sun, Hualei and Wang, Meng and others},
  journal={arXiv:2412.09375},
  year={2024},
  url={https://arxiv.org/abs/2412.09375}
}

@article{shi2025prerequisite,
  title={Prerequisite of superconductivity: SDW rather than tetragonal structure in double-layer {L}a$_3${N}i$_2${O}$_{7-x}$},
  author={Shi, Mengzhu and Peng, Di and Li, Yikang and Xing, Zhenfang and Wang, Yuzhu and Fan, Kaibao and Li, Houpu and Wu, Rongqi and Zeng, Zhidan and Zeng, Qiaoshi and others},
  journal={arXiv:2501.14202},
  year={2025},
  url={https://arxiv.org/abs/2501.14202}
}

@article{li2025ambient,
  title={Ambient pressure growth of bilayer nickelate single crystals with superconductivity over 90 K under high pressure},
  author={Li, Feiyu and Peng, Di and Dou, Jie and Guo, Ning and Ma, Liang and Liu, Chao and Wang, Lingzhen and Zhang, Yulin and Luo, Jun and Yang, Jie and others},
  journal={arXiv:2501.14584},
  year={2025},
  url={https://arxiv.org/abs/2501.14584}
}

@article{huo2025low,
  title={Low volume fraction of high-$T_c$ superconductivity in {L}a$_3${N}i$_2${O}$_{7}$ at 80 K and ambient pressure},
  author={Huo, Mengwu and Ma, Peiyue and Huang, Chaoxin and Huang, Xing and Sun, Hualei and Wang, Meng},
  journal={arXiv:2501.15929},
  year={2025},
  url={https://arxiv.org/abs/2501.15929}
}

@article{zhang2025damage,
  title={Damage of bilayer structure in {L}a$_3${N}i$_2${O}$_{7-\delta}$ induced by high pO2 annealing},
  author={Zhang, Yulin and Pei, Cuiying and Guo, Ning and Fan, Longlong and Zhang, Mingxin and Wang, Lingzhen and Zhang, Gongting and Li, Feiyu and Wang, Yunong and Ma, Chao and others},
  journal={arXiv:2502.01501},
  year={2025},
  url={https://arxiv.org/abs/2502.01501}
}

@article{YaoDX2023,
  title = {Bilayer Two-Orbital Model of {L}a$_3${N}i$_2${O}$_7$ under Pressure},
  author = {Luo, Zhihui and Hu, Xunwu and Wang, Meng and W\'u, W\'ei and Yao, Dao-Xin},
  journal = {Phys. Rev. Lett.},
  volume = {131},
  issue = {12},
  pages = {126001},
  numpages = {6},
  year = {2023},
  month = {Sep},
  publisher = {American Physical Society},
  doi = {10.1103/PhysRevLett.131.126001},
  url = {https://link.aps.org/doi/10.1103/PhysRevLett.131.126001}
}

@article{lu2023bilayertJ,
  title = {Interlayer-Coupling-Driven High-Temperature Superconductivity in {L}a$_3${N}i$_2${O}$_7$ under Pressure},
  author = {Lu, Chen and Pan, Zhiming and Yang, Fan and Wu, Congjun},
  journal = {Phys. Rev. Lett.},
  volume = {132},
  issue = {14},
  pages = {146002},
  numpages = {6},
  year = {2024},
  month = {Apr},
  publisher = {American Physical Society},
  doi = {10.1103/PhysRevLett.132.146002},
  url = {https://link.aps.org/doi/10.1103/PhysRevLett.132.146002}
}

@article{oh2023type2,
  title = {Type-{II} $t$-${J}$ model and shared superexchange coupling from Hund's rule in superconducting {L}a$_3${N}i$_2${O}$_7$},
  author = {Oh, Hanbit and Zhang, Ya-Hui},
  journal = {Phys. Rev. B},
  volume = {108},
  issue = {17},
  pages = {174511},
  numpages = {8},
  year = {2023},
  month = {Nov},
  publisher = {American Physical Society},
  doi = {10.1103/PhysRevB.108.174511},
  url = {https://link.aps.org/doi/10.1103/PhysRevB.108.174511}
}

@article{zhang2023structural,
  title={Structural phase transition, $s_{\pm}$-wave pairing, and magnetic stripe order in bilayered superconductor {L}a$_3${N}i$_2${O}$_7$ under pressure},
  author={Zhang, Yang and Lin, Ling-Fang and Moreo, Adriana and Maier, Thomas A and Dagotto, Elbio},
  journal={Nat. Commun.},
  volume={15},
  number={1},
  pages={2470},
  year={2024},
  publisher={Nature Publishing Group UK London},
  url={https://www.nature.com/articles/s41467-024-46622-z}
}

@article{qin2023high,
  title={High-$T_c$ superconductivity by mobilizing local spin singlets and possible route to higher T c in pressurized {L}a$_3${N}i$_2${O}$_7$},
  author={Qin, Qiong and Yang, Yi-feng},
  journal={Phys. Rev. B},
  volume={108},
  number={14},
  pages={L140504},
  year={2023},
  publisher={APS},
  url={https://journals.aps.org/prb/abstract/10.1103/PhysRevB.108.L140504}
}

@article{qu2023roles,
  title={Roles of Hund's Rule and Hybridization in the Two-orbital Model for High-$T_c$ Superconductivity in the Bilayer Nickelate},
  author={Qu, Xing-Zhou and Qu, Dai-Wei and Li, Wei and Su, Gang},
  journal={arXiv:2311.12769},
  year={2023},
  url={https://arxiv.org/abs/2311.12769}
}

@article{liao2023electron,
  title = {Electron correlations and superconductivity in {L}a$_{3}${N}i$_{2}${O}$_{7}$ under pressure tuning},
  author = {Liao, Zhiguang and Chen, Lei and Duan, Guijing and Wang, Yiming and Liu, Changle and Yu, Rong and Si, Qimiao},
  journal = {Phys. Rev. B},
  volume = {108},
  issue = {21},
  pages = {214522},
  numpages = {9},
  year = {2023},
  month = {Dec},
  publisher = {American Physical Society},
  doi = {10.1103/PhysRevB.108.214522},
  url = {https://link.aps.org/doi/10.1103/PhysRevB.108.214522}
}

@article{HuJP2023,
  title={Effective model and pairing tendency in bilayer {N}i-based superconductor {L}a$_3${N}i$_2${O}$_7$},
  author={Gu, Yuhao and Le, Congcong and Yang, Zhesen and Wu, Xianxin and Hu, Jiangping},
  journal={arXiv:2306.07275},
  url = {https://arxiv.org/abs/2306.07275},
  year={2023}
}


\noindent{\bf Acknowledgments}

\noindent We are grateful to the discussions with Yu-Bo Liu and Hongyi Sun. This work is supported by the NSFC under Grant Nos.12234016, 12074031, 12141402, 12334002 and 12304180. Ming Zhang is supported by Zhejiang Provincial Natural Science Foundation of China under Grant No. ZCLQN25A0402. This work was supported by the CAS Superconducting Research Project under Grant No. [SCZX-0101].


\noindent{\bf Author contribution}

\noindent Fan Yang designed the project. Chen Lu and Ming Zhang carried out numerical calculations for SBMF calculations. Ming Zhang, Chen Lu, Zhiming Pan and Fan Yang wrote the manuscript. Congjun Wu and Fan Yang contributed to the project design and provided valuable insights. All co-authors provided useful comments and discussion on the paper.

~~~~~~~~~~~~~~~~~~~~~~~~~~~~~~~~~~~~~~~~~~~~~~~~~~~~~~~~~~~~~~~~~~~~~~~~~~~~~~~~~~~~~~~~~~~~~~~~~~~~~~~~~~~~~~~~~~~~~~~~~~~~~~~~~~~~~~~~~~~~~~~

~~~~~~~~~~~~~~~~~~~~~~~~~~~~~~~~~~~~~~~~~~~~~~~~~~~~~~~~~~~~~~~~~~~~~~~~~~~~~~~~~~~~~~~~~~~~~~~~~~~~~~~~~~~~~~~~~~~~~~~~~~~~~~~~~~~~~~~~~~~~~~~

\noindent{\bf Competing interests}

\noindent Fan Yang is an Editorial Board Member for Communications Physics, but was not involved in the editorial review of, or the decision to publish this article. All other authors declare no competing interests.

~~~~~~~~~~~~~~~~~~~~~~~~~~~~~~~~~~~~~~~~~~~~~~~~~~~~~~~~~~~~~~~~~~~~~~~~~~~~~~~~~~~~~~~~~~~~~~~~~~~~~~~~~~~~~~~~~~~~~~~~~~~~~~~~~~~~~~~~~~~~~~~

~~~~~~~~~~~~~~~~~~~~~~~~~~~~~~~~~~~~~~~~~~~~~~~~~~~~~~~~~~~~~~~~~~~~~~~~~~~~~~~~~~~~~~~~~~~~~~~~~~~~~~~~~~~~~~~~~~~~~~~~~~~~~~~~~~~~~~~~~~~~~~~

\noindent{\bf Addition information}

\noindent{\bf Supplementary information} The online version contains supplementary information available at
\end{document}